\documentclass[%
 reprint,
 superscriptaddress,
 amsmath,amssymb,
 aps,
]{revtex4-2}

\usepackage{graphicx}
\usepackage{dcolumn}
\usepackage{bm}

\usepackage{comment}
\usepackage{tikz}
\usepackage{color}
\usepackage{physics}
\usepackage{ulem}
\newcommand{\Jperp}{J_\perp}

\begin{document}

\preprint{APS/123-QED}

\title{
Continuous phase transition between N\'eel and valence bond solid phases \\
in a $J$-$Q$-like spin ladder system} 

\author{Takuhiro Ogino}
\affiliation{Institute for Solid State Physics, University of Tokyo, Kashiwa, Chiba 277-8581, Japan}
\author{Ryui Kaneko}
\affiliation{Department of Physics, Kindai University, Higashi-Osaka, Osaka 577-8502, Japan}
\affiliation{Institute for Solid State Physics, University of Tokyo, Kashiwa, Chiba 277-8581, Japan}
\author{Satoshi Morita}
\affiliation{Institute for Solid State Physics, University of Tokyo, Kashiwa, Chiba 277-8581, Japan}
\author{Shunsuke Furukawa}
\affiliation{Department of Physics, Keio University, Kohoku-ku, Yokohama, Kanagawa 223-8522, Japan}
\author{Naoki Kawashima}
\affiliation{Institute for Solid State Physics, University of Tokyo, Kashiwa, Chiba 277-8581, Japan}

\date{\today}

\begin{abstract}
We investigate a quantum phase transition 
between a N\'{e}el phase and a valence bond solid (VBS) phase, 
in each of which a different $\mathbb{Z}_2$ symmetry is broken, 
in a spin-$1/2$ two-leg XXZ ladder with a four-spin interaction. 
The model can be viewed as a one-dimensional variant of the celebrated $J$-$Q$ model on a square lattice.  
By means of variational uniform matrix product state calculations and an effective field theory, 
we determine the phase diagram of the model, and present evidences that the N\'eel-VBS transition is continuous and belongs to the Gaussian universality class with the central charge $c=1$. 
In particular, the critical exponents $\beta, \eta,$ and, $\nu$ are found to satisfy the constraints expected for a Gaussian transition within numerical accuracy.
These exponents do not detectably change along the phase boundary while they are in general allowed to do so for the Gaussian class.

\end{abstract}

\maketitle


\section{ Introduction \label{sec:Intro} }
Continuous transitions do not occur between two phases in each of which 
a different symmetry is broken spontaneously,
according to the Landau-Ginzburg-Wilson (LGW) paradigm. 
Contrary to this conventional wisdom, 
Senthil \textit{et al.}\ proposed a deconfined quantum critical point (DQCP),
at which a direct continuous transition between a N\'{e}el phase and a valence bond solid (VBS) phase
occurs in a two-dimensional (2D) spin-$1/2$ system~\cite{Senthil1490,PhysRevB.70.144407}.
The 2D DQCP is characterized by deconfined spin-$1/2$ spinons coupled to
an emergent U(1) gauge field, which results in the same critical exponents
for the order parameters in the N\'{e}el 
and VBS phases.
An unusually large anomalous dimension $\eta$ is also expected
because the spinons coupled to the emergent gauge field are not at all free particles.

One of the simplest microscopic models that 
may
realize deconfined quantum criticality (DQC) is the $J$-$Q$ model
on a 2D square lattice~\cite{PhysRevLett.98.227202}.
The model consists of the nearest-neighbor Heisenberg interaction
$J$ and a four-spin interaction $Q$,
and exhibits a 
phase transition between a N\'{e}el phase and a VBS phase.
A quantum Monte Calro (QMC) study suggests
a continuous nature of the N\'eel-VBS transition,
nearly the same critical exponents for the magnetic and VBS order parameters,
the unusually large anomalous dimension $\eta = 0.26\pm0.03$,
and the emergent U(1) symmetry, consistent with DQC
scenario~\cite{Senthil1490,PhysRevB.70.144407}.
On the other hand, the possibility of 
a weak first-order transition,
where the correlation length would exceed an accessible system size,
cannot be ruled out~\cite{jiang2008antiferromagnet,PhysRevLett.101.050405}.
Later, larger systems and higher symmetry (SU$(N)$ with $N=2,3,4$) 
were explored~\cite{PhysRevB.88.220408},
which revealed unusually strong corrections to scaling 
that make 
the estimate of critical indices systematically drift even beyond $L=256$, 
while the finite-size-scaling plot seemed reasonably good for restricted size windows.
These observations clearly shows computational difficulty in diagnosing a DQC in two spatial dimensions.

Recently, DQC in one spatial dimension has been intensively studied 
\cite{PhysRevB.99.075103,PhysRevB.99.165143,PhysRevB.99.205153,PhysRevB.100.064427,PhysRevB.100.121111,
PhysRevB.100.125137,PhysRevB.98.014414,roberts2020onedimensional}.
This is because 
analytical approaches such as bosonization and a slave particle theory 
may help understand the nature of DQC, especially in one dimension.
Besides, numerical methods based on matrix product states 
(MPS) 
are applicable to a 
rich 
variety of models including frustrated ones in one dimension.
We note that a continuous symmetry breaking is prohibited in 1D quantum systems 
unless the uniform magnetic susceptibility diverges \cite{Momoi1996}.
Therefore we 
have to introduce, 
for example, easy-axis anisotropy
or long-range exchange interactions to realize a magnetic long-range order.
The former examples include
an anisotropic XZ model with nearest- and next-nearest-neighbor
interactions~\cite{PhysRevB.99.165143,PhysRevB.99.075103}.
The model has a ferromagnetic (FM) phase and a VBS phase,
in each of which a different $\mathbb{Z}_2$ symmetry
is broken. 
The transition between the two phases was found to be continuous and to be 
characterized by the central charge $c=1$.
The critical exponents in the FM phase are very close to those in the VBS phase
and change continuously along the phase boundary.
The latter examples include 
the $J$-$Q$ chain with long-range Heisenberg 
interactions~\cite{yang2020deconfined}.
The model exhibits 
an AFM-VBS transition, which takes place 
between two gapless phases.

In this paper, we investigate a quantum phase transition between two ordered phases, in each of which a different $\mathbb{Z}_2$ symmetry is broken, 
in a $J$-$Q$-like model on a two-leg ladder. 
In contrast to the 
FM-VBS transition
studied in Ref.\ \cite{PhysRevB.99.165143,PhysRevB.99.075103}, 
we propose the model which realizes a N\'{e}el phase and the VBS phase,
as the original 2D $J$-$Q$ model does. 
Our model consists of short-range interactions up to four neighboring sites. 
We introduce easy-axis XXZ anisotropy 
to realize the N\'{e}el phase.
The four-spin interaction $Q$ on each plaquette of a square lattice reduces to that on each plaquette of a two-leg ladder. 
For simplicity, we omit the 
four-spin
interaction between rungs
and keep only that between 
the 
legs, which we call $J_4$.
The $J_4$ interaction introduces effective repulsion between singlet pairs on each plaquette, and induces a staggered dimer (SD) phase, which is a type of VBS phase.
Analyzing this model will 
provide a step toward understanding 
the relationship 
between the N\'{e}el-VBS transitions 
in one dimension~\cite{PhysRevB.25.4925,PhysRevB.36.5291,kuboki1987spin,nomura1993phase,nomura1994critical,PhysRevB.99.205153,PhysRevB.81.094430}
and 
those 
in two dimensions~\cite{Senthil1490,PhysRevB.70.144407,PhysRevLett.98.227202}.

We obtain the ground-state phase diagram of
the spin-$1/2$ XXZ model on a two-leg ladder with the four-spin interaction
by means of numerical calculations and an effective field theory. 
We obtain a rung singlet (RS) phase, the N\'eel phase, and 
the SD phase 
as the four-spin interaction $J_4$ is increased.
The effective field theory for weak inter-chain couplings, 
which is based on bosonization, 
suggests that 
the RS-N\'eel transition 
belongs to the 2D Ising universality class, and 
the N\'eel-SD transition belongs 
to the Gaussian universality class.
However, this observation is based on leading terms in the effective Hamiltonian, and 
it is not obvious whether possible perturbations can modify the scenario. 
Furthermore, the effective field theory may fail when the inter-chain couplings are comparable to the intra-chain coupling. 
For these reasons, we numerically study the N\'{e}el-SD transition.
Our numerical calculations are based on the variational uniform matrix product state algorithm (VUMPS) \cite{PhysRevB.97.045145,10.21468/SciPostPhysLectNotes.7}, 
by which the ground state of the translationally invariant infinite system can be obtained in the form of an MPS.
Our numerical results support 
the scenario 
that the N\'{e}el-SD (VBS) transition in the 1D $J$-$Q$-like model belongs to the Gaussian universality class.

We comment on an important difference between our model and the original $J$-$Q$ model studied in Ref.\ \cite{PhysRevLett.98.227202}: 
the 
four-spin interaction has the positive coefficient $J_4>0$ in the former 
and the negative coefficient $-Q<0$ in the latter. 
Therefore, the original $J$-$Q$ model exhibits a columnar dimer (CD) phase instead of the SD phase. 
With a positive coefficient $-Q>0$, the $J$-$Q$ model on a square lattice is also expected to exhibit the SD phase, 
which has indeed been found in a closely related model with ring exchange \cite{PhysRevLett.95.137206}. 
However, detailed properties of the N\'eel-SD transition has yet to be explored on a square lattice 
because the QMC suffers from a sign problem for $-Q>0$. 
In contrast, with a negative coefficient $J_4<0$, our model on a ladder exhibits a transition between the RS and CD phases, 
and the N\'eel phase does not appear 
between them 
(this can be understood from the field-theoretical analysis in Sec.\ \ref{sec:EFT}). 
This suggests that the models with a positive coefficient in the 
four-spin interaction are more suitable for studying the relationship between the N\'eel-VBS transitions in one and two dimensions. 
This observation, together with indications of the continuous N\'eel-SD (VBS) transition on a ladder in the present study, would stimulate a numerical study on the $J$-$Q$ model with $-Q>0$ on a square lattice using, e.g., tensor network algorithms \cite{ORUS2014117,PhysRevLett.101.250602,PhysRevLett.101.090603}.

We organize this paper as follows.
In Sec.\ \ref{sec:Model}, we introduce our model 
and briefly review the previous related studies. 
In Sec.\ \ref{sec:EFT}, we present a field-theoretical analysis for weak inter-chain couplings. 
In Sec.\ \ref{sec:Results},
we present a detailed numerical analysis on the transition between the  N\'{e}el phase and the VBS (SD) phase. 
In Sec.\ \ref{sec:conclusion}, we draw our conclusion.

\section{ Model \label{sec:Model} }

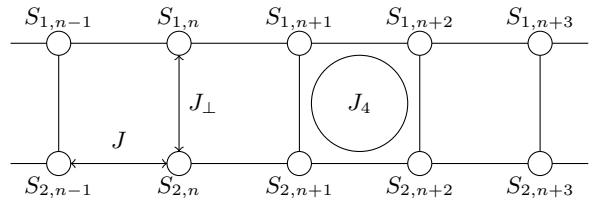
\begin{figure}[b]
\begin{center}
\vspace{1cm}
\begin{tikzpicture}[scale=0.8]
\draw (-0.8,0) -- (-0.2,0);\draw (-0.8,2) -- (-0.2,2);
\node at (0,-0.4) {$S_{2,n-1}$}; \node at (0,2.4) {$S_{1,n-1}$};
\draw  (0,0) circle (0.2cm);\draw (0,2) circle (0.2cm); \draw (0,0.2) -- (0,1.8);
\draw [<->] (0.2,0) -- (1.8,0);\draw (0.2,2) -- (1.8,2);
\node at (2,-0.4) {$S_{2,n}$}; \node at (2,2.4) {$S_{1,n}$};
\draw (2,0) circle (0.2cm);\draw (2,2) circle (0.2cm);\draw [<->](2,0.2) -- (2,1.8);
\draw (2.2,0) -- (3.8,0);\draw (2.2,2) -- (3.8,2);
\node at (4,-0.4) {$S_{2,n+1}$}; \node at (4,2.4) {$S_{1,n+1}$};
\draw (4,0) circle (0.2cm);\draw (4,2) circle (0.2cm);\draw (4,0.2) -- (4,1.8);
\draw (4.2,0) -- (5.8,0);\draw (4.2,2) -- (5.8,2);
\node at (6,-0.4) {$S_{2,n+2}$}; \node at (6,2.4) {$S_{1,n+2}$};
\draw (6,0) circle (0.2cm);\draw (6,2) circle (0.2cm);\draw (6,0.2) -- (6,1.8);
\draw (6.2,0) -- (7.8,0);\draw (6.2,2) -- (7.8,2);
\node at (8,-0.4) {$S_{2,n+3}$}; \node at (8,2.4) {$S_{1,n+3}$};
\draw (8,0) circle (0.2cm);\draw  (8,2) circle (0.2cm);\draw (8,0.2) -- (8,1.8);
\draw (8.2,0) -- (8.8,0);\draw (8.2,2) -- (8.8,2);
\node at (1,0.4) {$J$}; \node at (2.4,1) {$J_\perp$}; \node at (5,1) {$J_4$};\draw (5,1) circle (0.8cm);
\end{tikzpicture}
\caption{Two-leg XXZ spin ladder system with a four-spin interaction, 
which is described by the Hamiltonian \eqref{eq:modelmain}.
}
\label{fig.system}
\end{center}
\end{figure}

\begin{figure}[b]

\begin{center}

\begin{tikzpicture}[scale=0.5]
\node at (-2.0,1) {$(a)$};
\draw (-0.8,0) -- (-0.2,0);\draw (-0.8,2) -- (-0.2,2);
\draw (0,0) circle (0.2cm);\draw (0,2) circle (0.2cm); \draw (0,0.2) -- (0,1.8);
\draw (0.2,0) -- (1.8,0);\draw (0.2,2) -- (1.8,2);
\draw (2,0) circle (0.2cm);\draw  (2,2) circle (0.2cm);\draw (2,0.2) -- (2,1.8);
\draw (2.2,0) -- (3.8,0);\draw (2.2,2) -- (3.8,2);
\draw (4,0) circle (0.2cm);\draw (4,2) circle (0.2cm);\draw (4,0.2) -- (4,1.8);
\draw (4.2,0) -- (5.8,0);\draw (4.2,2) -- (5.8,2);
\draw (6,0) circle (0.2cm);\draw  (6,2) circle (0.2cm);\draw (6,0.2) -- (6,1.8);
\draw (6.2,0) -- (7.8,0);\draw (6.2,2) -- (7.8,2);
\draw (8,0) circle (0.2cm);\draw  (8,2) circle (0.2cm);\draw (8,0.2) -- (8,1.8);
\draw (8.2,0) -- (8.8,0);\draw (8.2,2) -- (8.8,2);
\draw [dashed](0,1) circle (0.5cm and 1.5cm); \draw [dashed](2,1) circle (0.5cm and 1.5cm); \draw [dashed](4,1) circle (0.5cm and 1.5cm);
\draw [dashed](6,1) circle (0.5cm and 1.5cm); \draw [dashed](8,1) circle (0.5cm and 1.5cm);
\end{tikzpicture}

\begin{tikzpicture}[scale=0.5]
\node at (-2.0,1) {$(b)$};
\node at (0,3) {$$};
\draw (-0.8,0) -- (-0.2,0);\draw (-0.8,2) -- (-0.2,2);
\draw (0,0) circle (0.2cm);\draw (0,2) circle (0.2cm); \draw (0,0.2) -- (0,1.8);
\draw (0.2,0) -- (1.8,0);\draw (0.2,2) -- (1.8,2);
\draw (2,0) circle (0.2cm);\draw  (2,2) circle (0.2cm);\draw (2,0.2) -- (2,1.8);
\draw (2.2,0) -- (3.8,0);\draw (2.2,2) -- (3.8,2);
\draw (4,0) circle (0.2cm);\draw (4,2) circle (0.2cm);\draw (4,0.2) -- (4,1.8);
\draw (4.2,0) -- (5.8,0);\draw (4.2,2) -- (5.8,2);
\draw (6,0) circle (0.2cm);\draw  (6,2) circle (0.2cm);\draw (6,0.2) -- (6,1.8);
\draw (6.2,0) -- (7.8,0);\draw (6.2,2) -- (7.8,2);
\draw (8,0) circle (0.2cm);\draw  (8,2) circle (0.2cm);\draw (8,0.2) -- (8,1.8);
\draw (8.2,0) -- (8.8,0);\draw (8.2,2) -- (8.8,2);
\draw [dashed](1,0) circle (1.5cm and 0.5cm); \draw [dashed](3,2) circle (1.5cm and 0.5cm); \draw [dashed](5,0) circle (1.5cm and 0.5cm);
\draw [dashed](7,2) circle (1.5cm and 0.5cm);
\end{tikzpicture}

\begin{tikzpicture}[scale=0.5]
\node at (-2.0,1) {$(c)$};
\node at (0,3) {$$};
\draw (-0.8,0) -- (-0.2,0);\draw (-0.8,2) -- (-0.2,2);
\draw (0,0) circle (0.2cm);\draw (0,2) circle (0.2cm); \draw (0,0.2) -- (0,1.8);
\draw (0.2,0) -- (1.8,0);\draw (0.2,2) -- (1.8,2);
\draw (2,0) circle (0.2cm);\draw  (2,2) circle (0.2cm);\draw (2,0.2) -- (2,1.8);
\draw (2.2,0) -- (3.8,0);\draw (2.2,2) -- (3.8,2);
\draw (4,0) circle (0.2cm);\draw (4,2) circle (0.2cm);\draw (4,0.2) -- (4,1.8);
\draw (4.2,0) -- (5.8,0);\draw (4.2,2) -- (5.8,2);
\draw (6,0) circle (0.2cm);\draw  (6,2) circle (0.2cm);\draw (6,0.2) -- (6,1.8);
\draw (6.2,0) -- (7.8,0);\draw (6.2,2) -- (7.8,2);
\draw (8,0) circle (0.2cm);\draw  (8,2) circle (0.2cm);\draw (8,0.2) -- (8,1.8);
\draw (8.2,0) -- (8.8,0);\draw (8.2,2) -- (8.8,2);
\node at (0,0) {\rotatebox{20}{\huge$\uparrow$}}; 
\node at (0,2) {\rotatebox{20}{\huge$\downarrow$}}; 
\node at (2,2) {\rotatebox{20}{\huge$\uparrow$}}; 
\node at (2,0) {\rotatebox{20}{\huge$\downarrow$}}; 
\node at (4,0) {\rotatebox{20}{\huge$\uparrow$}}; 
\node at (4,2) {\rotatebox{20}{\huge$\downarrow$}}; 
\node at (6,2) {\rotatebox{20}{\huge$\uparrow$}}; 
\node at (6,0) {\rotatebox{20}{\huge$\downarrow$}}; 
\node at (8,0) {\rotatebox{20}{\huge$\uparrow$}}; 
\node at (8,2) {\rotatebox{20}{\huge$\downarrow$}}; 
\end{tikzpicture}

\caption{
(a) The rung singlet (RS) phase, (b) the staggered dimer (SD) phase, and (c) the N\'eel phase. 
Two spins enclosed by a dashed line form a singlet. 
The SD phase spontaneously breaks a $\mathbb{Z}_2$ symmetry with respect to the rung-centered reflection. 
The N\'eel phase has an antiferromagnetic order along the $z$-axis, 
and spontaneously breaks a $\mathbb{Z}_2$ symmetry with respect to the global $\pi$ rotation of spins about the $x$-axis. 
In this phase, the ordered moment $\langle S_{\alpha,i}^z \rangle$ shows a staggered pattern along both the leg and rung directions. 
}
\label{fig:phases}
\end{center}
\end{figure}
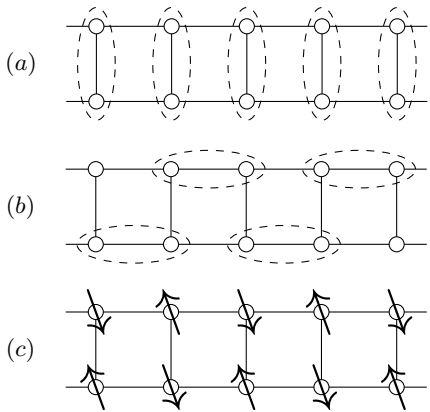


We study a spin-$1/2$ ladder model which is described by the Hamiltonian
\begin{equation}\label{eq:modelmain}
\begin{split}
H  &:= J \sum_{\alpha = 1,2} \sum_{j}  (\bm{S}_{\alpha,j}\cdot\bm{S}_{\alpha,j+1})_\Delta
	 + \Jperp \sum_j (\bm{S}_{1,j}\cdot\bm{S}_{2,j})_\Delta \\
   &+ J_4\sum_{j}(\bm{S}_{1,j}\cdot\bm{S}_{1,j+1})(\bm{S}_{2,j}\cdot\bm{S}_{2,j+1})
\end{split}
\end{equation}
with
\begin{equation}\label{eq:model}
 (\bm{S}_{\alpha,i}\cdot\bm{S}_{\beta,j})_\Delta := S^x_{\alpha,i} S^x_{\beta,j} + S^y_{\alpha,i} S^y_{\beta,j} + \Delta S^z_{\alpha,i} S^z_{\beta,j} .
\end{equation}
Here, the $J$ and $\Jperp$ terms represent the nearest-neighbor XXZ interactions with the anisotropy parameter $\Delta$ along the legs and the rungs, respectively,
and the $J_4$ term represents the four-spin interactions (Fig.\ \ref{fig.system}).
Throughout this paper, we take the leg interaction $J=1$ as the unit of energy, and focus on the case of  $\Jperp,J_4\ge 0$ and $\Delta$ close to unity.

When $J_4=0$ and $\Delta=1$, the model \eqref{eq:modelmain} is a standard antiferromagnetic Heisenberg model on a ladder. 
For $\Jperp>0$, it shows a rung singlet (RS) phase [Fig. \ref{fig:phases} (a)]\cite{PhysRevLett.69.2419,PhysRevB.50.9911,PhysRevB.53.8521}, 
which has a unique ground state below an excitation gap. 
The case in which the anisotropy $\Delta$ is introduced for the leg interaction has also been investigated by field-theoretical and numerical approaches \cite{PhysRevLett.69.2419,PhysRevB.50.9911,PhysRevB.72.014449}. 
In the proposed phase diagram, the N\'eel phase with an antiferromagnetic order along the $z$-axis  [Fig. \ref{fig:phases} (c)] appears for easy-axis anisotropy $\Delta>1$ and the sufficiently weak antiferromagnetic inter-chain coupling $\Jperp>0$. 
The N\'eel phase has two degenerate ground states below an excitation gap, and is characterized by the order parameter 
\begin{equation}\label{eq:ONeel}
\expval{\mathcal{O}_{\text{N\'{e}el}}(j) }
= \frac{1}{4} 
\expval{ S_{1,j}^{z} -  S_{2,j}^{z} - S_{1,j+1}^{z} + S_{2,j+1}^{z} } .
\end{equation}
In this phase, a $\mathbb{Z}_2$ symmetry with respect to the global $\pi$ rotation of spins about the $x$-axis ($S_{\alpha,j}^{y} \mapsto -S_{\alpha,j}^{y}$ and $S_{\alpha,j}^{z} \mapsto -S_{\alpha,j}^{z}$) is spontaneously broken. 

The case of $\Jperp=0$ and $\Delta=1$ has been studied as a 1D spin-orbital model. 
At $J_4=4$, the model has an enhanced SU$(4)$ symmetry, and is solvable by the Bethe ansatz \cite{PhysRevLett.81.3527}. 
At this point, the system is gapless and is described effectively by the SU$(4)_1$ Wess-Zumino-Witten (WZW) model with the central charge $c=3$ \cite{affleck1986exact,affleck1988critical}. 
Through numerical and field-theoretical analyses \cite{PhysRevLett.81.5406,PhysRevB.61.6747}, it has been found that a gapless phase continues for $J_4\ge 4$ 
while a staggered dimer (SD) phase  [Fig. \ref{fig:phases} (b)] with two degenerate ground states below an excitation gap appears for $0<J_4<4$. 
The SD phase is characterized by the order parameter 
\begin{eqnarray}\label{eq:OSD}
\expval{\mathcal{O}_{\text{SD}}(j) }
&=& \frac{1}{4} 
\big\langle \bm{S}_{1,j-1}\cdot\bm{S}_{1,j} - \bm{S}_{2,j-1}\cdot\bm{S}_{2,j} \nonumber \\
&& - \bm{S}_{1,j}\cdot\bm{S}_{1,j+1} + \bm{S}_{2,j}\cdot\bm{S}_{2,j+1} \big\rangle.
\end{eqnarray}
In this phase, a $\mathbb{Z}_2$ symmetry with respect to the rung-centered reflection (${\bm S}_{\alpha,j} \mapsto {\bm S}_{\alpha,-j}$) is spontaneously broken. 
Intuitively, the SD order is a consequence of effective repulsion between singlet pairs in the same plaquette due to $J_4$. 

When both $\Jperp>0$ and $J_4>0$ are present in the isotropic case $\Delta=1$, the RS and SD phases compete. 
The boundary between these phases in the $\Jperp$-$J_4$ plane has been obtained numerically for $0<\Jperp,J_4\lesssim 2$ \cite{PhysRevB.80.014426, PhysRevLett.122.027201} 
(see Refs.\ \cite{PhysRevB.67.100409,PhysRevLett.90.087204,PhysRevB.65.104413,PhysRevB.68.134403} for related studies on a spin ladder model with ring exchange). 
Field-theoretical analyses for weak inter-chain couplings \cite{PhysRevLett.78.3939,PhysRevB.66.134423,PhysRevB.82.214420,PhysRevLett.122.027201} suggest 
that this transition is continuous and is described by the SU$(2)_2$ WZW theory (equivalent to three copies of free massless Majorana fields) with the central charge $c=3/2$. 
The exact diagonalization result of Ref.\ \cite{PhysRevB.80.014426} is consistent with this scenario for $0.5\lesssim \Jperp \lesssim 1.5$. 
We note that the phase diagram is expected to have a more complex structure for larger $J_4$. 
In fact, for $J_4=4$, the exact solution of Ref.\ \cite{PhysRevLett.81.3527} can be extended for $\Jperp\ne 0$ \cite{PhysRevB.60.9236}. 
It shows two gapless phases 
over $\Jperp < -\pi/2\sqrt{3} + (\ln 3)/2(\simeq -0.36)$ and $-\pi/2\sqrt{3} + (\ln 3)/2 < \Jperp < 4$ 
as well as the gapped RS phase for $\Jperp > 4$. 

In this paper, we consider the case in which $\Delta$ is slightly larger than unity, i.e., in a weakly easy-axis regime. 
We find that there appears a finite region of the N\'eel phase that intervenes between the RS and SD phases (see Fig.\ \ref{fig:phasediagram} presented later). 
This N\'eel phase is expected to be adiabatically connected to the one discussed for the XXZ ladder in Refs.\ \cite{PhysRevLett.69.2419,PhysRevB.50.9911,PhysRevB.72.014449}. 
It is remarkable that just the addition of small Ising interactions ($\Delta-1>0$) changes the phase structure. 
This phase structure can be derived in the field-theoretical analysis for weak inter-chain couplings $0\le \Jperp,J_4\ll 1$ (Sec.\ \ref{sec:EFT}), 
and is numerically confirmed for $\Jperp=1$ (Sec.\ \ref{sec:Results}). 
Our particular interest lies in the nature of the transition between the N\'eel and SD phases, each of which spontaneously breaks a different $\mathbb{Z}_2$ symmetry. 
Using the field-theoretical and numerical approaches, we will argue that a Gaussian transition with the central charge $c=1$ is the most plausible scenario for this transition at least in the parameter range of our interest.

\section{ Effective field theory for weak inter-chain couplings \label{sec:EFT} }
\subsection{Bosonization }

For weak inter-chain couplings with $|\Jperp|,|J_4| \ll 1$, the ground-state phase diagram of the model \eqref{eq:modelmain} can be studied 
by means of effective field theory based on bosonization \cite{giamarchi2003quantum,gogolin2004bosonization}. 
Our formulation is an extension of those in Refs.\ \cite{PhysRevLett.69.2419,PhysRevB.50.9911,PhysRevB.53.8521,PhysRevLett.78.3939,PhysRevB.82.214420,PhysRevLett.122.027201}, 
and we take similar notations as those in Ref.\ \cite{PhysRevB.82.214420}.
Our starting point is the two decoupled XXZ chains obtained for $\Jperp=J_4=0$. 
In this case, each chain labeled by $\alpha=1,2$ is described effectively by the quantum sine-Gordon Hamiltonian
\begin{equation}\label{eq:sineGordon}
\begin{split}
 H_\alpha^\mathrm{eff} = &\int \mathrm{d}x \frac{v}{2} \left[K^{-1} \left(\partial_x\phi_\alpha\right)^2+ K \left(\partial_x\theta_\alpha\right)^2 \right] \\
 &- \frac{v\lambda}{2\pi} \cos(4\sqrt{\pi}\phi_\alpha).
\end{split}
\end{equation}
Here, the bosonic field  $\phi_\alpha (x)$ and its dual counterpart $\theta_\alpha(x)$ satisfy the commutation relation 
$[\phi_\alpha(x),\theta_\beta(x')] = -i\delta_{\alpha\beta}\vartheta_{\text{step}}(x-x')$ with $\vartheta_{\text{step}}(\cdot)$ being the Heaviside step function. 
For $\Delta\le 1$, the $\lambda$ term is irrelevant (marginally irrelevant at $\Delta=1$), 
and thus the effective Hamiltonian in the infrared limit is given by the Gaussian part of Eq.\ \eqref{eq:sineGordon}, which is known as the Tomonaga-Luttinger liquid (TLL) theory. 
In this limit, the spin velocity $v$ and the TLL parameter $K$ can be obtained exactly from the Bethe ansatz. 
The TLL parameter $K$ monotonically decreases as a function of $\Delta$, and reaches $K=1/2$ at $\Delta=1$. 
When $\Delta$ exceeds unity, the $\lambda$ term with the scaling dimension $4K$ becomes relevant (i.e., $4K<2$), and grows along the renormalization group (RG) flow. 
Owing to $\lambda>0$, this term eventually leads to the locking of the bosonic field at $2\sqrt{\pi}\phi_\alpha=0,\pi$, 
which correspond to the N\'eel order in the $z$ direction [as seen in Eq.\ \eqref{eq:Sz_bos} below]. 
In this case, $K$ and $\lambda$ depend on the scale of our concern. 
We note that for $\Delta>1$ and with an infinitesimal antiferromagnetic inter-chain coupling $\Jperp>0$, the N\'eel states on the individual chains are interlocked, 
leading to the N\'eel state on a ladder (Fig.\ \ref{fig:phases} (c)) as discussed in Refs.\ \cite{PhysRevLett.69.2419,PhysRevB.50.9911,PhysRevB.72.014449}. 

The spin operators on each chain are related to the bosonic fields as 
\begin{subequations}\label{eq:SzSpm_bos}
\begin{align}
S_{\alpha, j}^z &= \frac{a}{\sqrt{\pi}} \partial_x \phi_\alpha + (-1)^j a_1 \cos(2\sqrt{\pi} \phi_\alpha) + \cdots, \label{eq:Sz_bos}\\
S_{\alpha,j}^+ &= e^{i\sqrt{\pi}\theta_\alpha} \qty[b_0 (-1)^j + b_1 \cos(2\sqrt{\pi}\phi_\alpha) + \cdots],
\end{align}
\end{subequations}
where the fields are taken at $x=ja$ with $a$ being the lattice constant. 
Furthermore, the dimer operators, i.e., the product of neighboring spins, are also related to the fields as 
\begin{subequations}\label{eq:dim_bos}
\begin{align}
&  S_{\alpha,j}^z S_{\alpha,j+1}^z  = (-1)^j d_{z} \sin(2\sqrt{\pi}\phi_\alpha) + \cdots,\\
& S_{\alpha,j}^x S_{\alpha, j+1}^x + S_{\alpha,j}^y S_{\alpha,j+1}^y = (-1)^j d_{xy} \sin(2\sqrt{\pi}\phi_\alpha) + \cdots, 
\end{align}
\end{subequations}
where the fields are taken at $x=(j+1/2)a$ and the uniform component is omitted for simplicity. 
The non-universal coefficients $a_1$, $b_0$, $b_1$, $d_z$, and $d_{xy}$ in Eqs.\ \eqref{eq:SzSpm_bos} and \eqref{eq:dim_bos} have been determined analytically and numerically 
for $\Delta< 1$ \cite{LUKYANOV1997571,PhysRevB.58.R583,PhysRevB.82.214420,PhysRevB.96.134429}. 
For $\Delta=1$, these coefficients depend on the scale of our concern because of the presence of the marginally irrelevant perturbation;  
at a fixed length scale, they should satisfy $a_1 = b_0 =: \bar{a}$ and $d_z=d_{xy}/2 =: \bar{d}$ because of the SU$(2)$ symmetry. 

Let us now include the inter-chain couplings, $J_\perp$ and $J_4$, perturbatively. 
Using Eqs.\ \eqref{eq:SzSpm_bos} and \eqref{eq:dim_bos} for the inter-chain couplings, 
the low-energy effective Hamiltonian of Eq.~(\ref{eq:modelmain}) is obtained as 
\begin{equation}\label{eq:Heff}
\begin{split}
H^\mathrm{eff}=
& \int \dd x \sum_{q=\pm} \frac{v_q}{2} \qty[K_q^{-1}\qty(\partial_x\phi_q) + K_q \qty(\partial_x\theta_q)^2] \\
&+ g_+ \cos(2\sqrt{2\pi}\phi_+) + g_- \cos(2\sqrt{2\pi}\phi_-) \\
&+ \tilde{g}_- \cos(\sqrt{2\pi}\theta_-) + \cdots.
\end{split}
\end{equation}
Here, we introduced symmetric and antisymmetric combinations of the fields, $\phi_{\pm} := (\phi_1 \pm \phi_2)/\sqrt{2}$ and $\theta_{\pm} := (\theta_1 \pm \theta_2)/\sqrt{2}$. 
The coupling constants in Eq.\ \eqref{eq:Heff} are given in terms of the inter-chain couplings as 
\begin{equation}\label{eq:coeff_cos}
 g_\pm = \frac1a \qty(J_{\perp} \Delta \frac{a_1^2}{2} \mp J_4 \frac{(3d)^2}{2}),~~
 \tilde{g}_-=\frac{1}{a} J_\perp b_0^2,
\end{equation}
where $3d:=d_{xy}+d_z$. 
Furthermore, the velocities $v_\pm$ and the TLL parameters $K_\pm$ are now defined for the symmetric and antisymmetric channels, 
and they are in general modified from the values $v$ and $K$ in the decoupled XXZ chains by the effects of the inter-chain couplings. 
In Eq.\ \eqref{eq:Heff}, we focused on terms that are most important around the SU$(2)$-symmetric case $\Delta=1$. 
Indeed, the $g_\pm$ and $\tilde{g}_-$ terms have the scaling dimensions $2K_\pm$ and $1/(2K_-)$, respectively, 
which are all equal to unity in the limit of the decoupled Heisenberg chains. 
These terms are much more relevant, in the RG sense, than the $\lambda$ term with the scaling dimension $2K_++2K_-$ in Eq.\ \eqref{eq:sineGordon}. 
Therefore, the $\lambda$ term can be ignored unless the anisotropy $\Delta-1$ is significantly larger than $J_\perp$ and $J_4$. 

\subsection{Expected phase diagram}\label{sec:bos_phases}

The effective Hamiltonian \eqref{eq:Heff} indicates a separation of the symmetric and antisymmetric channels. 
The symmetric channel is described by the sine-Gordon model, 
in which the strongly relevant $g_+$ term leads to the locking of $\phi_+$ at distinct positions depending on the sign of $g_+$. 
A Gaussian-type transition with the central charge $c=1$ is expected at $g_+=0$. 
The antisymmetric channel is described by the dual-field double sine-Gordon model, in which the strongly relevant $g_- $ and $\tilde{g}_-$ terms compete. 
When $K_-=1/2$, both the terms have the same scaling dimensions of unity, 
and the long-distance physics can be determined by simply examining which of $|g_-|$ and $|\tilde{g}_-|$ is larger 
(in this case, the model is known as the self-dual sine-Gordon model \cite{LECHEMINANT2002502}). 
Namely, $|g_-|>|\tilde{g}_-|$ ($|g_-|<|\tilde{g}_-|$) leads to the locking of $\phi_-$ ($\theta_-$). 
In fact, the self-dual sine-Gordon model can be mapped onto two free Majorana fields---one of them is always massive (unless $g_-=\tilde{g}_-=0$)
while the other is massless at  $|g_-|=|\tilde{g}_-|$ and massive otherwise \cite{PhysRevB.53.8521,LECHEMINANT2002502}. 
Therefore, an Ising-type transition with the central charge $c=1/2$ is expected at $|g_-|=|\tilde{g}_-|$, which corresponds to a free massless Majorana field. 
When $K_-$ deviates slightly from $1/2$, a similar picture is still expected to hold as the change in $K_-$ is a marginal perturbation. 
In the present argument, we have ignored perturbations which have larger scaling dimensions than the $g_\pm$ and $\tilde{g}$ terms in Eq.\ \eqref{eq:Heff}. 
If such terms also become relevant, they can potentially change the nature of the phase transitions. 
We will address this issue later. 

We are ready to discuss the expected phase diagram of the model \eqref{eq:modelmain}. 
We assume that $\Delta$ is slightly larger than unity, i.e., in a weakly easy-axis regime. 
We fix the values of $J_\perp>0$ and $\Delta>1$, and vary $J_4\ge 0$. 
We then find two phase transitions as follows. 
The first transition occurs at $J_{4,c}^{\text{Ising}} \approx \qty[ (2b_0^2-\Delta a_1^2)/(3d)^2 ]J_\perp$, 
which is an Ising-type transition with the central charge $c=1/2$ in the antisymmetric channel. 
The second transition occurs at $J_{4,c}^{\text{Gauss}} = \Delta \qty(a_1/3d)^2J_\perp$, 
which is a Gaussian-type transition with the central charge $c=1$ in the symmetric channel. 
For $J_4< J_{4,c}^{\text{Ising}}$, the coupling constants in Eq.\ \eqref{eq:coeff_cos} satisfy $g_+>0$ and $0<g_-<\tilde{g}_-$, 
and the resulting state is characterized by the field locking at 
\begin{equation}\label{eq:RS_bos}
 (2\sqrt{2\pi}\phi_+,\sqrt{2\pi}\theta_-)=(\pi,\pi).
\end{equation}
This corresponds to the RS phase, which is known in an antiferromagnetic ladder model \cite{PhysRevLett.69.2419,PhysRevB.50.9911,PhysRevB.53.8521}.  
For $J_{4,c}^{\text{Ising}}<J_4<J_{4,c}^{\text{Gauss}}$, $g_-$ becomes larger than $\tilde{g}_-$, 
resulting in the field locking at 
\begin{equation}\label{eq:Neel_bos}
 2\sqrt{2\pi} (\phi_+, \phi_-)=(\pi,\mp \pi).
\end{equation} 
These correspond to the N\'eel phase with 
\begin{equation}\label{eq:ONeel_bos}
\begin{split}
 \expval{\mathcal{O}_{\text{N\'{e}el}}(j) }
 &= -  (-1)^j a_1 \expval{ \sin \qty(\sqrt{2\pi}\phi_+) \sin \qty(\sqrt{2\pi}\phi_-) }\\
 &=\pm (-1)^j c_\mathrm{N},
\end{split}
\end{equation}
where $\mathcal{O}_{\text{N\'{e}el}}(j)$ is defined in Eq.\ \eqref{eq:ONeel} and $c_\mathrm{N}>0$ is a constant independent of $j$. 
For $J_4>J_{4,c}^{\text{Gauss}}$, $g_+$ becomes negative, resulting in the field locking at 
\begin{equation}\label{eq:SD_bos}
 2\sqrt{2\pi} (\phi_+,\phi_-)=(0,\mp \pi).
\end{equation} 
These correspond to the SD phase with 
\begin{equation}\label{eq:OSD_bos}
\begin{split}
 \expval{\mathcal{O}_{\text{SD}}(j) } 
 &= -(-1)^j (3d) \expval{ \cos \qty(\sqrt{2\pi}\phi_+) \sin \qty(\sqrt{2\pi}\phi_-) }\\
 &=\pm (-1)^j c_\mathrm{SD},
\end{split}
\end{equation}
where $\mathcal{O}_{\text{SD}}(j)$ is defined in Eq.\ \eqref{eq:OSD} and $c_\mathrm{SD}>0$ is again a constant independent of $j$. 
In the isotropic limit $\Delta\to 1$, the two transition points $J_{4,c}^{\text{Ising}}$ and $J_{4,c}^{\text{Gauss}}$ 
merge into the single point $J_{4,c}=  \qty(\bar{a}/3\bar{d})^2J_\perp$, 
at which the central charge is expected to be $c=3/2$ \cite{PhysRevB.80.014426,PhysRevB.82.214420,PhysRevLett.78.3939,PhysRevLett.122.027201,PhysRevB.66.245106,PhysRevB.66.134423}. 
In Ref.\ \cite{PhysRevB.82.214420}, this point was estimated to be $J_{4,c}=2.05J_\perp$ using the numerical values of $\bar{a}$ and $\bar{d}$ in the Heisenberg chain at a certain scale. 
More detailed phase diagrams in the $\Jperp$-$J_4$ plane in the isotropic case $\Delta=1$ have been obtained numerically in Refs.\ \cite{PhysRevB.80.014426,PhysRevLett.122.027201}. 

In this paper, we are particularly interested in the nature of the transition between the N\'eel and SD phases. 
While the effective Hamiltonian \eqref{eq:Heff} suggests that this transition is likely to be of Gaussian type, 
we have to examine whether possible perturbations to the theory can modify this scenario. 
Since the antisymmetric channel remains gapped at this transition, we can focus on the symmetric channel. 
As a possible perturbation, we can consider, for example, 
a higher-frequency cosine potential $\cos\left( 4\sqrt{2\pi}\phi_+ \right)$ with the scaling dimension $8K_+$. 
If this term becomes relevant, it can crucially change the nature of the phase transition \cite{PhysRevB.81.094430}. 
With a negative coefficient, for example, this term has minima at $2\sqrt{2\pi}\phi_+=0$ and $\pi$, 
which correspond to the N\'eel and SD orders in Eqs.\ \eqref{eq:Neel_bos} and \eqref{eq:SD_bos}. 
Different signs of $g_+$ select different types of orders, and thus the first-order transition at $g_+=0$ separates the two phases. 
In our numerical results for $\Jperp=1$ in Sec.\ \ref{sec:Results}, 
$K_+$ is estimated to be $K_+\approx 0.6$ as shown in Table \ref{tab:exp},
and thus the above higher-frequency cosine term is expected to be irrelevant. 

In passing, we comment on the case of $J_\perp>0$ and $J_4<0$, 
which has a closer form to the original $J$-$Q$ model on a square lattice studied in Ref.\ \cite{PhysRevLett.98.227202}. 
In this case, the symmetric channel remains gapped as $g_+$ is always positive. 
An Ising transition in the antisymmetric channel occurs at $\tilde{J}_{4,c}^{\text{Ising}} \approx -\qty[ (\Delta a_1^2+2b_0^2)/(3d)^2 ]J_\perp$ \cite{PhysRevB.82.214420}. 
For $J_4>\tilde{J}_{4,c}^{\text{Ising}}$, we have $|g_-|<\tilde{g}_-$, which results in the RS phase with Eq.\ \eqref{eq:RS_bos}. 
For $J_4<\tilde{J}_{4,c}^{\text{Ising}}$, we have $g_-<-\tilde{g}_-<0$, which results in the field locking at
\begin{equation}\label{eq:CD_bos}
 2\sqrt{2\pi} (\phi_+,\phi_-)=(\pi, 2\pi),~(\pi,0).
\end{equation} 
These correspond to the CD phase with 
\begin{equation}\label{eq:OCD_bos}
\begin{split}
 \expval{\mathcal{O}_{\text{CD}}(j) } 
 &= \frac{1}{4} 
\big\langle \bm{S}_{1,j-1}\cdot\bm{S}_{1,j} + \bm{S}_{2,j-1}\cdot\bm{S}_{2,j} \nonumber \\
 &~~~ - \bm{S}_{1,j}\cdot\bm{S}_{1,j+1} - \bm{S}_{2,j}\cdot\bm{S}_{2,j+1} \big\rangle\\
 &= -(-1)^j (3d) \expval{ \sin \qty(\sqrt{2\pi}\phi_+) \cos \qty(\sqrt{2\pi}\phi_-) }\\
 &=\pm (-1)^j c_\mathrm{CD},
\end{split}
\end{equation}
where $c_\mathrm{CD}>0$ is a constant independent of $j$. 
We therefore have the RS-CD transition of an Ising type, which is robust against the introduction of the XXZ anisotropy. 
This is why we focus on the region of $J_\perp>0$ and $J_4>0$ in our study of the N\'eel-VBS transition.

\subsection{Critical properties around the Gaussian transition}\label{sec:bos_critical}

Assuming that the N\'eel-SD transition is of Gaussian type, 
we discuss the scaling behavior of physical quantities around this transition. 
We first consider the correlation functions of the N\'eel and dimer operators, $\mathcal{O}_{\text{N\'{e}el}}(j)$ and $\mathcal{O}_{\text{SD}}(j)$, at the transition point. 
As seen in the bosonized expressions in Eqs.\ \eqref{eq:ONeel_bos} and \eqref{eq:OSD_bos}, these operators involve both the fields $\phi_\pm$. 
As the symmetric channel is described by the Gaussian theory at the transition point, 
the symmetric component $\sin\qty(\sqrt{2\pi}\phi_+)$ or $\cos\qty(\sqrt{2\pi}\phi_+)$ shows a critical correlation with the decay exponent $K_+$. 
In contrast, as $\sqrt{2\pi}\phi_-$ remains locked at $\mp \pi/2$, the antisymmetric component $\sin\qty(\sqrt{2\pi}\phi_-)$ shows a correlation 
that converges to a non-vanishing constant above a certain length scale proportional to the inverse of the excitation gap. 
Therefore, above this scale, the correlation functions in total exhibit the power-law behavior
\begin{subequations}\label{eq:corr_N_SD}
\begin{align}
 C_{\text{N\'{e}el}}(r)
 &:= (-1)^r \expval{ \mathcal{O}_{\text{N\'{e}el}}(r) \mathcal{O}_{\text{N\'{e}el}}(0) } \sim r^{-K_+},\label{eq:corr_N} \\
 C_{\text{SD}}(r)
 &:= (-1)^r \expval{ \mathcal{O}_{\text{SD}}(r) \mathcal{O}_{\text{SD}}(0) } \sim r^{-K_+}. \label{eq:corr_SD} 
\end{align}
\end{subequations}

We next discuss the scaling behavior when the system deviates slightly from the critical point $J_{4,c}^\mathrm{Gauss}$. 
In this case, the dimensionless coupling constant $G_+:=g_+ a^2/(2\pi v)\propto (J_4-J_{4,c})/J$ grows along the RG flow as $G(\ell)=G_+ e^{(2-2K_+)\ell}$ as the short-distance cutoff $a$ is changed to $e^\ell a$. 
We continue the RG transformation until the running coupling constant $G_+ (\ell)$ becomes $O(1)$. 
We suppose that the correlation length $\xi$ is a constant $\xi_0$ in units of $e^\ell a$ for this $\ell$. 
We then find
\begin{equation}
 \xi \sim \xi_0 e^\ell a \sim |G_+|^{-\nu} \sim \qty| J_4-J_{4,c}^\mathrm{Gauss} |^{-\nu}
\end{equation}
with the exponent
\begin{equation}\label{eq:nu_Kp}
 \nu=\frac{1}{2-2K_+}.
\end{equation}
In the N\'eel phase ($J_4<J_{4,c}^\mathrm{Gauss}$), the correlation function $C_{\text{N\'{e}el}}(r)$ 
shows the power-law behavior \eqref{eq:corr_N}  below the scale of $\xi$, 
and converges to a non-vanishing constant above this scale. 
This means that the N\'eel order parameter acquires a nonzero expectation value
\begin{align}
 (-1)^j \expval{\mathcal{O}_{\text{N\'{e}el}}(j)} \sim  \pm \sqrt{\xi^{-K_+}} \sim \pm \qty( J_{4,c}^\mathrm{Gauss}-J_4 )^\beta
\end{align} 
with the exponent 
\begin{equation}\label{eq:beta_Kp}
 \beta=\frac{\nu K_+}{2} = \frac{K_+}{4-4K_+}.
\end{equation}
Likewise, in the SD phase  ($J_4>J_{4,c}^\mathrm{Gauss}$), the dimer order parameter acquires a nonzero expectation value
\begin{equation}
 (-1)^j \expval{ \mathcal{O}_{\text{SD}}(j) } \sim \pm \qty( J_4-J_{4,c}^\mathrm{Gauss} )^\beta
\end{equation}
with the same exponent $\beta$ as above.

\section{ Numerical analysis of the N\'{e}el-SD transition \label{sec:Results} }
\subsection{ Method }
We have performed numerical calculations directly for the infinite system 
by using the VUMPS algorithm \cite{PhysRevB.97.045145,10.21468/SciPostPhysLectNotes.7}. 
In this algorithm, a variational state is prepared in the form of a uniform MPS by assuming the translational invariance, and the ground state is obtained by iteratively optimizing the constituent tensors to lower the variational energy. 

The original VUMPS is an algorithm for
1D chains. 
It can be applied to the present ladder system by regarding two sites on each rung as a single effective site with the local Hilbert space dimension of four. 
We have adopted the two-site unit cell implementation in Ref.\ \cite{PhysRevB.97.045145} 
as all the phases discussed in Sec.\ \ref{sec:Model} and Sec.\ \ref{sec:EFT} have the unit cell consisting of at most two effective sites (i.e., two rungs). 
We omit data points that do not converge 
sufficiently 
and use the data points that have the gradient norm $\|B\| < 10^{-10}$,
where $\|B\|$ is defined as the gradient of energy per site with respect to
the elementary tensor in VUMPS.

\subsection{ 
Correlation length and entanglement entropy
\label{subsec:NSD}} 
\begin{figure}[t]
\includegraphics[width=90mm]{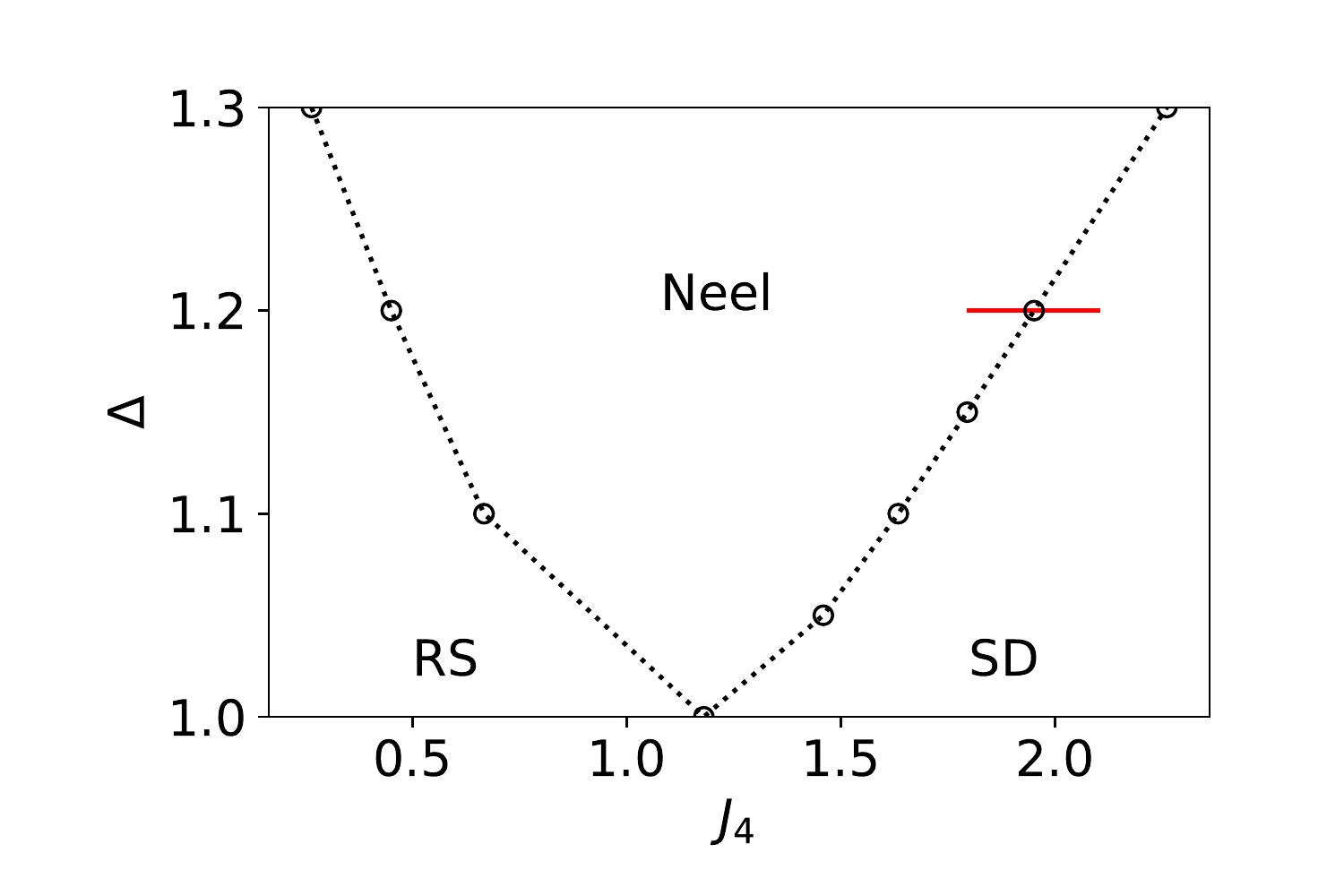}
\caption{
Phase diagram of the model (\ref{eq:modelmain}) on the $J_4$-$\Delta$ plane with $J=J_\perp=1$. 
The circles ($\circ$) indicate the transition points obtained through the analysis of the correlation length 
as shown in Figs.\ \ref{fig:corrchi} and \ref{fig:CC}.
The dotted lines are our assumption of the phase boundaries.
We will focus on the phase transition along the red solid line with $\Delta = 1.2$ in Sec.\ \ref{sec:Results}.
The numerical analysis of the RS-N\'eel transition is described in Appendix \ref{subsec:RSNeel}. 
The transition point $J_{4,c}\simeq 1.19$ in the isotropic case $\Delta=1$ 
has been 
obtained by the exact diagonalization \cite{PhysRevB.80.014426}.
We also obtained the same critical point within the error bar of bond dimension extrapolation by VUMPS, 
as described in Appendix \ref{app:RSSD}. 
}
\label{fig:phasediagram}
\end{figure}

We fix $J=\Jperp=1$, and present the numerically obtained phase diagram in the $J_4$-$\Delta$ plane 
in Fig.\ \ref{fig:phasediagram}.
The N\'{e}el phase and the SD phase break different $\mathbb{Z}_2$ symmetries.
Below we study the N\'{e}el-SD transition along the red solid line ($\Delta = 1.2$)
shown in Fig. \ref{fig:phasediagram} as a representative case. 


\begin{figure}[b]
\includegraphics[width=90mm]{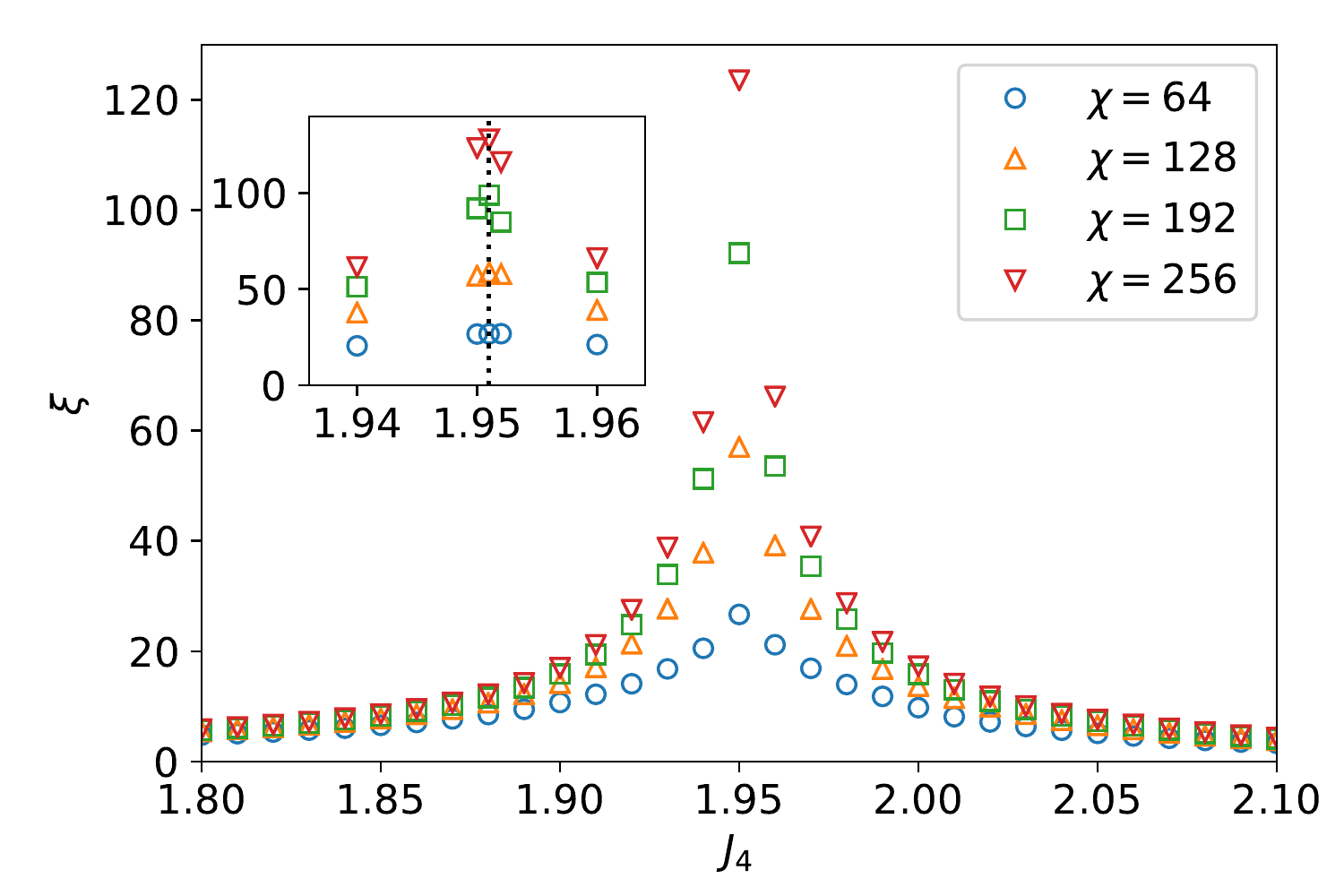}
\caption{
Correlation length 
$\xi$ as a function of $J_4$ around the N\'eel-SD transition for $\Delta = 1.2$ 
(see the red solid line in Fig.\ \ref{fig:phasediagram}).
The correlation length shows 
a sharp peak, 
which grows consistently with an increase in $\chi$.
In the case of $\chi=256$, 
the peak of 
the correlation length exceeds $100$ lattice spacings. 
These results are indicative of a continuous phase transition.
We estimate the critical point to be $J_{4,c} = 1.951(1)$ as shown in the inset. }
\label{fig:corrchi}
\end{figure}

We extract the correlation length $\xi$ from the MPS with the finite bond dimension $\chi$.
In the case of two-site unit cells, the transfer matrix $T$ is defined in the following way.
We cast two-site unit cells into the single-site unit cells
by introducing 
$\mathbb{A}^{s} := A(1)^{s_1}A(2)^{s_2}$,
where  $A(k)^{s_{k}} \in \mathbb{R}^{\chi \times \chi}$ is the original matrix for the state $s_k$ at the $k$-th effective site and $\mathbb{A}^{s} \in \mathbb{R}^{\chi \times \chi}$ is the combined matrix for the state $s=(s_1,s_2)$.
The transfer matrix $T$ is 
then 
defined by $T:= \sum_s \mathbb{A}^{s\dagger}\otimes\mathbb{A}^{s}$.
The correlation length is calculated as
$\xi(\chi) = - 2 / \ln|\lambda_2(\chi)|$,
where $\lambda_2(\chi)$ is the second largest 
absolute eigenvalue of the transfer matrix.
This method underestimates the correlation length 
when the bond dimension $\chi$ is finite.
We plot the correlation length $\xi(\chi)$ at $\Delta = 1.2$ in Fig.\ \ref{fig:corrchi}.
It shows that the correlation length has a sharp peak with consistent growth with an increase in $\chi$.
In particular, the peak of 
the correlation length $\xi(\chi)$ 
exceeds $100$ lattice spacings  
in the case of $\chi = 256$.
These results are indicative of a continuous phase transition 
with a diverging correlation length.

The critical point can be determined from the peak of the correlation length. 
We find that the correlation length is largest at $J_4=1.951$ as shown in Fig.~\ref{fig:corrchi}.
We therefore estimate the critical point to be $J_{4,c} = 1.951(1)$.

\begin{figure}[t]
\includegraphics[width=90mm]{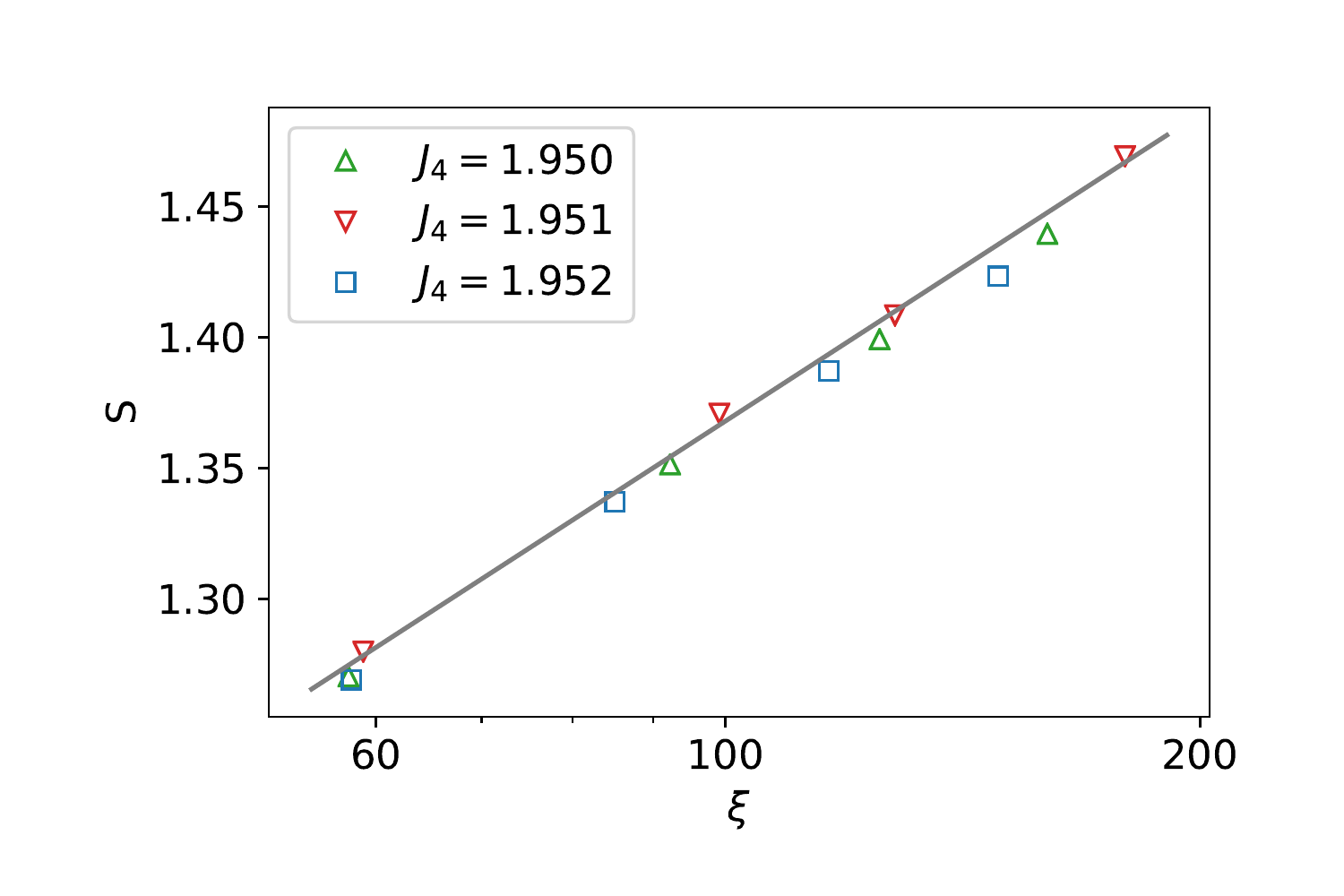}
\caption{
Entanglement entropy $S(\chi)$ versus the correlation length $\xi(\chi)$ for the bond dimensions $\chi=128,192,256,320$. 
These are calculated at three points near the N\'eel-SD transition point for $\Delta = 1.2$. 
A logarithmic scale is used for the horizontal axis.
The gray straight line 
shows the CFT formula \eqref{eq:cardy} 
with $c\simeq1.02$ and $S_0\simeq0.59$, which is 
the result of 
a
fitting to the case 
of 
$J_4 = 1.951$. 
\label{fig:CC}
}
\end{figure}

Critical points of a large class of 1D quantum
systems 
are described by the conformal field theory (CFT). 
A convenient quantity for probing the underlying CFT is the entanglement entropy $S$. 
We calculate it for a bipartition of the infinite 1D system into two half-infinite chains.
According to the CFT, 
the entanglement entropy $S$ and the correlation length $\xi$ have the relationship
\begin{eqnarray}
S = \frac{c}{6} \ln \xi + S_0,
\label{eq:cardy}
\end{eqnarray}
where $c$ is the central charge and $S_0$ is a constant \cite{Calabrese_2004,PhysRevLett.102.255701}.
Figure \ref{fig:CC}
shows that $S$ and $\xi$ at $J_4 = 1.951$ 
are well fitted by Eq.\ (\ref{eq:cardy}) with $c\simeq1.02$.
This result is also indicative of a continuous phase transition
because $S$ and $\xi$ do not follow Eq.\ (\ref{eq:cardy}) if the transition is a discontinuous one.
This result is consistent with 
the Gaussian transition with $c=1$ suggested by the effective field theory.


\begin{figure}[t]
\includegraphics[width=90mm]{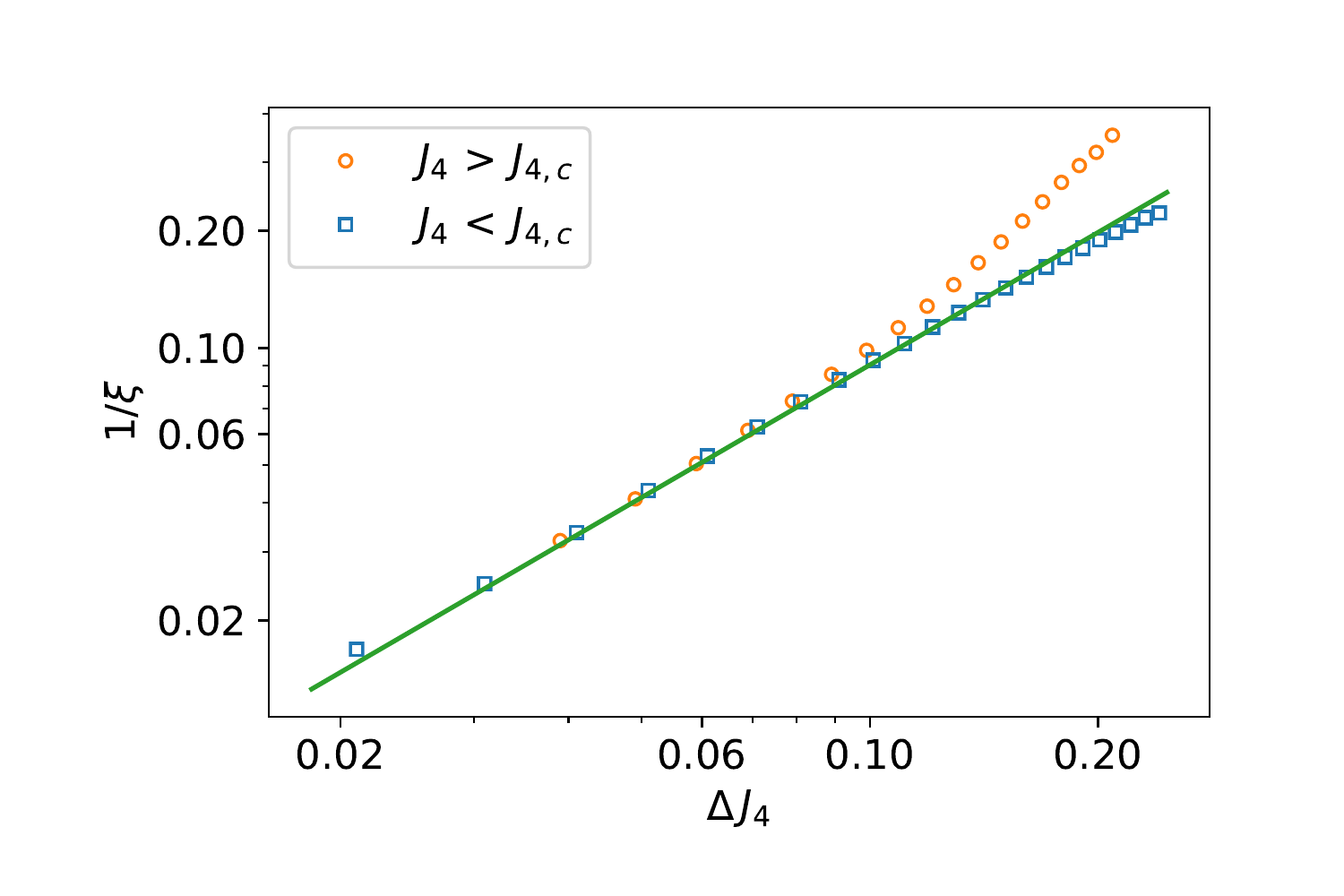}
\caption{
Power-law scaling behavior of the correlation length
$\xi$ as a function of $\Delta J_4 := |J_4 - J_{4,c}|$ around the N\'eel-SD transition point $J_{4,c} = 1.951$ for $\Delta = 1.2$. 
Logarithmic scales are used for both axes. 
For each $J_4$, the correlation length $\xi$ has been extrapolated to infinite $\chi$ as described in Appendix \ref{app:corr}, and the present figure shows the extrapolated values.
By using the region where the data points follow Eq.\ (\ref{eq:nu}), we obtain $\nu_{-} = 1.13(11)$ and $\nu_{+} = 1.26(16)$ in the N\'eel ($J_4<J_{4,c}$) and SD ($J_4>J_{4,c}$) phases, respectively.
The data in the two phases tend to collapse onto the same curve when approaching the critical point, which suggests the emergent symmetry.
The solid line is a guide to the eye and its slope is $1.13$.
}
\label{fig:nu}
\end{figure}

\begin{figure}[b]
\includegraphics[width=90mm]{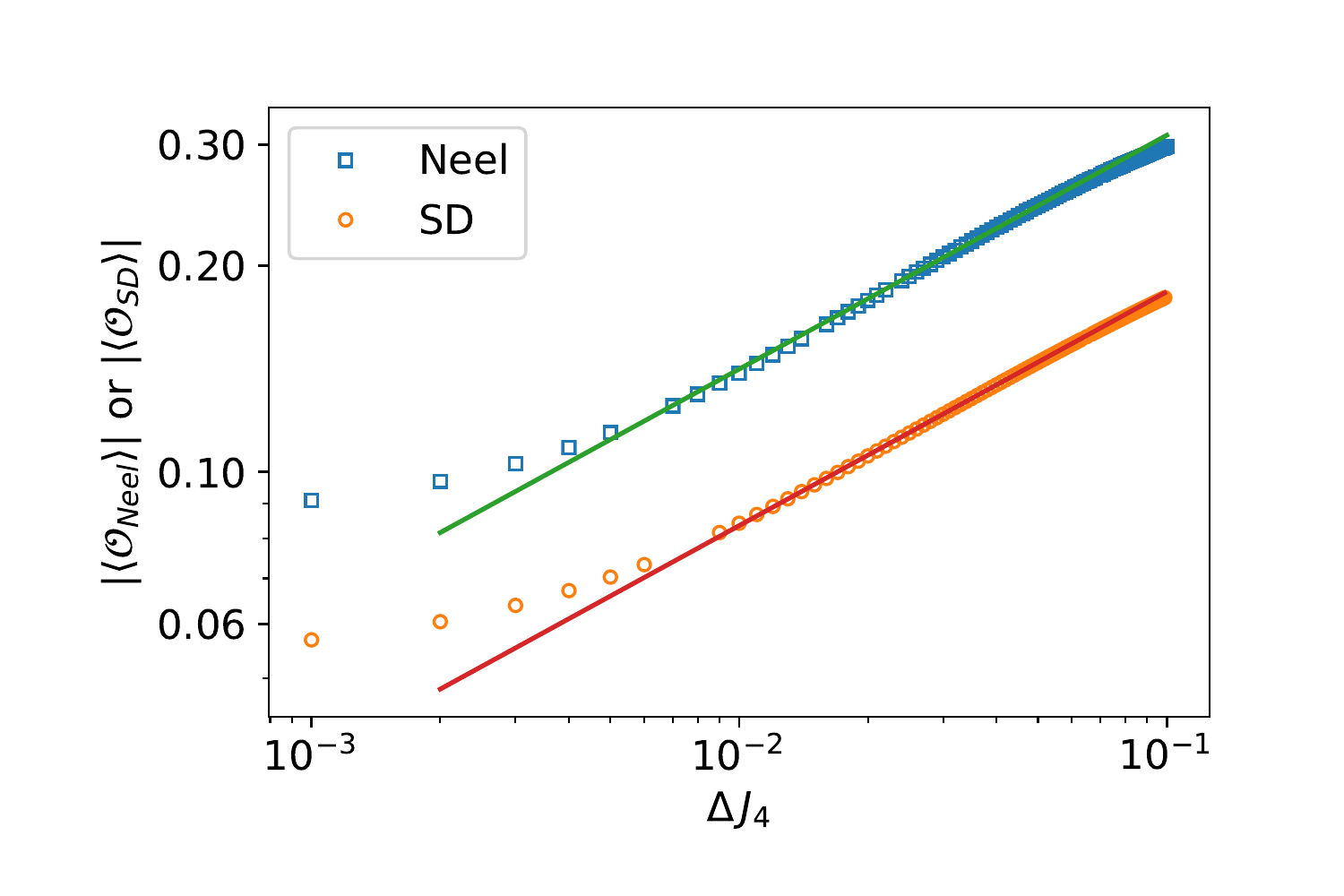}
\caption{
Power-law scaling behavior Eq.\ \eqref{eq:Neel_SD_scaling} of the N\'eel and SD order parameters [defined in Eqs.\ \eqref{eq:ONeel} and \eqref{eq:OSD}]  
around the transition point $J_{4,c} = 1.951$ for $\Delta = 1.2$. 
Logarithmic scales are used for both axes. 
The calculations are performed with the bond dimension $\chi=256$. 
The order parameter exponents 
are estimated to be 
$\beta_{\text{N\'{e}el}} = 0.34(3)$ and $\beta_{\text{SD}} = 0.34(3)$
as described in Appendix \ref{app:beta}. 
The two solid parallel lines are guides to the eye
and their slope is $0.34$.
}
\label{fig:beta}
\end{figure}

\subsection{Critical exponents}

We proceed to analyze the critical exponents around the estimated critical point $J_{4,c}$. 
First, we calculate the correlation length exponents $\nu_{\pm}$.
Near the critical point, the correlation length $\xi$ is expected to obey the scaling
\begin{eqnarray}
\xi = 
\begin{cases}
\xi_{-} (J_{4,c} - J_4 )^{-\nu_{-}} 
&(J_4<J_{4,c})\\
\xi_{+} (J_{4} - J_{4,c} )^{-\nu_{+}} 
&(J_4>J_{4,c}),
\end{cases}
\label{eq:nu}
\end{eqnarray}
where $\nu_\pm$ are the critical exponents and $\xi_\pm$ are the amplitudes in the N\'eel ($-$) and SD ($+$) phases. 
In the numerical result shown in Fig. \ref{fig:nu}, 
we find a
linear relation between
$\ln (1/\xi)$ 
and $\ln \Delta J_4$, where $\Delta J_4 := |J_4 - J_{4,c}|$. 
It also shows that when $\Delta J_4$ is large, the correlation length $\xi$ does not follow Eq.\ (\ref{eq:nu}).
By using the region where the data points follow Eq.\ (\ref{eq:nu}), we obtain $\nu_{-} = 1.13(11)$ and $\nu_{+} = 1.26(16)$. 
In these estimates, we have also taken account of a possible error in the estimate of the transition point $J_{4,c}$ as described in Appendix \ref{app:corr}. 
In Fig.\ \ref{fig:nu}, we observe that when approaching the transition point, the correlation length $\xi$ in the N\'eel and SD phases tends to collapse onto the same curve. 
This suggests that both the exponents $\nu_\pm$ and the amplitudes $\xi_\pm$ are equal between the two sides, indicating the emergent symmetry at the critical point \cite{PhysRevX.7.031051,PhysRevX.7.041016}. 
While $\nu_\pm$ obtained above agree with each other within the estimated error, due to the relatively large error,
we cannot draw a definitive conclusion on the emergent symmetry scenario only from this result.

Second, we calculate the order parameter critical exponents $\beta_{\text{N\'{e}el}}$ and $\beta_{\text{SD}}$.
Near the critical point, the order parameters defined in Eqs.\ \eqref{eq:ONeel} and \eqref{eq:OSD} are expected to obey the scaling 
\begin{eqnarray}\label{eq:Neel_SD_scaling}
&&(-1)^j \langle \mathcal{O}_{\text{N\'{e}el/SD}} (j) \rangle \nonumber\\
~ &&=
\begin{cases}
\pm 
A_{\text{N\'{e}el/SD}}  ~ (\Delta J_4)^{\beta_{\text{N\'{e}el/SD}} } &(\text{N\'eel/SD phase})\\
0 
&(\text{otherwise}).
\end{cases}
\label{eq:beta}
\end{eqnarray}
where $A_{\text{N\'{e}el/SD}}$ are constants.
Numerical data of the order parameters are shown in Fig.\ \ref{fig:beta}, where we find a linear relation between $\ln |\langle \mathcal{O}_{\text{N\'{e}el/SD}} \rangle|$ and $\ln (\Delta J_4)$. 
We also find 
that when $\Delta J_4$ is large or small, the order parameters do not follow Eq.\ (\ref{eq:beta}).
By using the region where the data points follow Eq.\ \eqref{eq:beta},
we obtain $\beta_{\text{N\'{e}el}} = 0.34(3)$ and $\beta_{\text{SD}} = 0.34(3)$. 
This indicates the equal exponents between the N\'eel and SD phases in consistency with the emergent symmetry scenario. 
The details of the calculation are described in Appendix \ref{app:beta}.

\begin{figure}[t]
\includegraphics[width=90mm]{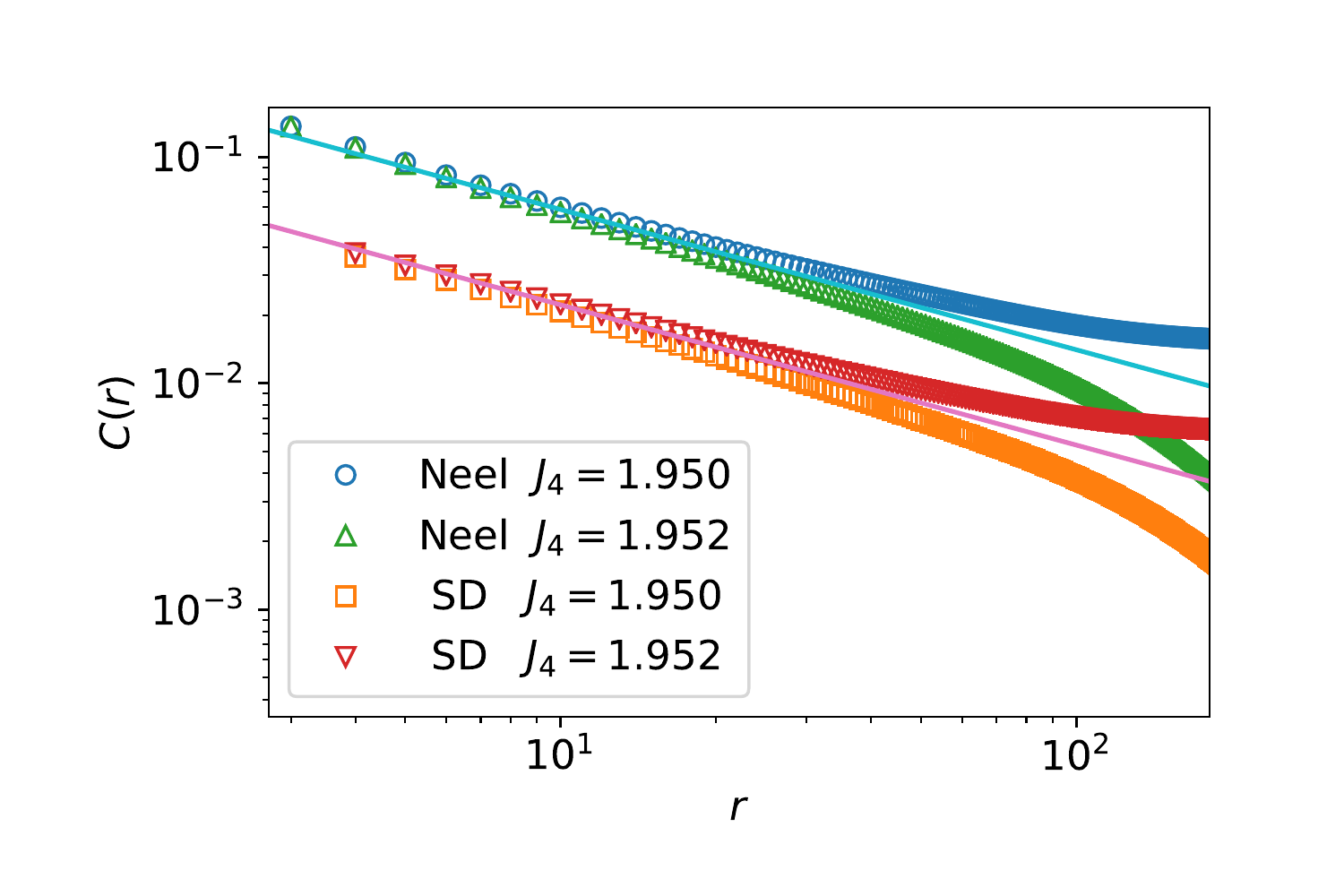}
\caption{
N\'eel and SD correlation functions at two points near the N\'eel-SD transition point $J_{4,c}=1.951(1)$ for $\Delta=1.2$. 
Logarithmic scales are used for both axes. 
The calculations are performed with $\chi=210$. 
The critical exponents are 
estimated to be 
$\eta_{\text{N\'{e}el}} = 0.63(6)$ and $\eta_{\text{SD}} = 0.61(5)$.
The two parallel solid lines are guides to the eye and 
their slope is $-0.62$.
}
\label{fig:corrfunc}
\end{figure}
\begin{table*}[t]
\caption{\label{tab:critical_exponents}
Numerical estimates of the 
critical point $J_{4,c}$ and the exponents 
$\beta_{\text{N\'eel/SD}}$, $\nu_\pm$, and $\eta_{\text{N\'eel/SD}}$ 
at several points along the N\'eel-SD phase boundary in Fig.\ \ref{fig:phasediagram}.
}
\label{tab:exp}
\begin{ruledtabular}
\begin{tabular}{llllllll}
\multicolumn{1}{c}{$\Delta$} & 
\multicolumn{1}{c}{$J_{4,c}$} & 
\multicolumn{1}{c}{$\beta_{\text{N\'{e}el}}$} &  
\multicolumn{1}{c}{$\beta_{\text{SD}}$}  & 
\multicolumn{1}{c}{$\nu_{-}$} & 
\multicolumn{1}{c}{$\nu_{+}$} & 
\multicolumn{1}{c}{$\eta_{\text{N\'{e}el}}$} & 
\multicolumn{1}{c}{$\eta_{\text{SD}}$} \\
\hline
$1.05$ & $1.459(1)$ & $0.35(2)$ &  $0.36(3)$  & $1.09(14)$ & $1.15(10)$ & $0.61(4)$ & 0.60(3) \\
$1.1$  & $1.634(1)$ & $0.35(2)$ &  $0.35(2)$  & $1.13(10)$ & $1.20(10)$ & $0.61(4)$ & 0.62(4) \\
$1.15$ & $1.795(1)$ & $0.34(2)$ &  $0.35(2)$ & $1.11(8)$  & $1.19(11)$ & $0.62(4)$ & 0.61(4) \\
$1.20$ & $1.951(1)$ & $0.34(3)$ &  $0.34(3)$ & $1.13(11)$ & $1.26(16)$ & $0.63(6)$ & 0.61(5) \\
$1.30$ & $2.261(1)$ & $0.34(3)$ &  $0.34(3)$ & $1.02(22)$ & $1.19(13)$ & $0.66(7)$ & 0.60(6) \\
\end{tabular}
\end{ruledtabular}
\end{table*}


Lastly, we
calculate the correlation functions in order to determine 
the exponents $\eta_{\text{N\'{e}el}}$ and $\eta_{\text{SD}}$.
As seen in Eq.\ \eqref{eq:corr_N_SD}, the correlation functions of the N\'eel and dimer operators are expected to show a power-law decay at the critical point:
\begin{equation}\label{eq:corr_scaling}
 C_{\text{N\'eel/SD}}(r) =C_{\text{N\'eel/SD}}^{(0)} r^{-\eta_{\text{N\'eel/SD}}},
\end{equation}
where $C_{\text{N\'eel/SD}}^{(0)}$ are constants. 
Figure \ref{fig:corrfunc} presents the correlation functions calculated for $J_4 = 1.950$ and $1.952$, which are close to the N\'eel-SD transition point. 
The N\'eel (SD) correlation function plotted in logarithmic scales is bent upward (downward) for $J_4=1.950$ and downward (upward) for $J_4=1.952$, 
indicating that the transition point is located inbetween them in accord with the analysis of the correlation length in Fig.\ \ref{fig:CC}. 
Although the two points $J_4 = 1.950$ and $1.952$ deviate slightly from the critical point, 
the correlation functions are expected to show a power-law behavior \eqref{eq:corr_scaling} below the scale of the correlation length. 
This can be confirmed via an approximately linear behavior at relatively short distances in Fig.\ \ref{fig:corrfunc}. 
The exponents $\eta_{\text{N\'eel/SD}}$ at the critical point should then exist between the slopes of the linear behaviors for $J_4 = 1.950$ and $1.952$. 
For the N\'eel correlation function, the slopes at $J_4 = 1.950$ and $1.952$ are estimated to be $-0.572$ and $-0.684$, respectively, using the region $ 7 < r < 34$. 
We therefore estimate the critical exponent as $\eta_{\text{N\'{e}el}} = 0.63(6)$. 
In the same way, we obtain $\eta_{\text{SD}} = 0.61(5)$.

As summarized in Table \ref{tab:critical_exponents},  
we have estimated the N\'eel-SD transition points and the critical exponents in a similar manner for some values of $\Delta$. 
We can confirm the relations $\beta_{\text{N\'eel}}=\beta_{\text{SD}}$, $\nu_-=\nu_+$, and $\eta_{\text{N\'eel}}=\eta_{\text{SD}}$ within numerical accuracy 
in consistency with the emergent symmetry scenario. 
We can further confirm the consistency with the constraints 
\begin{equation}\label{eq:nu_beta_eta}
 \beta=\frac{\eta}{4-4\eta},~~\nu=\frac{1}{2-2\eta},
\end{equation}
which are expected for the Gaussian universality class as discussed in Sec.\ \ref{sec:bos_critical} 
(here, the subscripts in the exponents are omitted assuming the emergent symmetry).  
For example, the substitution of $\eta_{\text{N\'eel}}=0.63(6)$ (the estimate for $\Delta=1.20$) into Eq.\ \eqref{eq:nu_beta_eta} gives $\beta=0.44(11)$ and $\nu=1.38(22)$, 
which are consistent with the estimates of $\beta_{\text{N\'eel/SD}}$ and $\nu_\pm$ in Table \ref{tab:critical_exponents} within numerical accuracy. 
Our numerical results thus support the scenario that the N\'eel-SD transition belongs to the Gaussian universality class. 
However, the exponents in Table \ref{tab:critical_exponents} do not detectably change along the phase boundary while they are in general allowed to do so for the Gaussian class. 
To detect possible changes in the exponent along the phase boundary, calculations with larger bond dimensions $\chi$ would be required.




\section{ Conclusion \label{sec:conclusion} }
In this paper, we 
have 
studied the spin-$1/2$ two-leg XXZ ladder system with a four-spin interaction,
which 
can be viewed as a 1D variant of the $J$-$Q$ model on a square lattice. 
We 
have 
determined the phase diagram and analyzed the nature of 
the quantum phase transitions 
by means of VUMPS calculations for the infinite system and an effective field theory based on bosonization. 
We
have presented evidences 
that the N\'{e}el-SD (VBS) transition 
belongs to 
the Gaussian universality class and
the RS-N\'{e}el transition
belongs to 
the 2D Ising universality class.

In particular, we have conducted detailed analyses of the N\'eel-SD transition, which occurs between two ordered phases breaking different $\mathbb{Z}_2$ symmetries.
The effective Hamiltonian [see Eq.\ \eqref{eq:Heff}]
for weak inter-chain couplings $|\Jperp|,|J_4| \ll J$  
indicates that this transition is described by a sine-Gordon model in the symmetric channel
and is likely to be of Gaussian type. 
However, it is not obvious whether possible perturbations to the effective Hamiltonian can modify the scenario or whether the effective Hamiltonian is applicable when $\Jperp$ and $J_4$ are comparable to $J$.
For these reasons, we numerically studied the N\'{e}el-SD transition.
The VUMPS study is consistent with the expected Gaussian universality class.
The correlation length shows a sharp peak 
that consistently grows with an increase in $\chi$. 
By using 
the entanglement entropy, 
it 
turned 
out that the transition has the expected central charge $c=1$.
The numerically estimated critical exponents in Table \ref{tab:critical_exponents} 
satisfy the relations $\beta_{\text{N\'eel}}=\beta_{\text{SD}}$, $\nu_-=\nu_+$, and $\eta_{\text{N\'eel}}=\eta_{\text{SD}}$ within numerical accuracy in consistency with the emergent symmetry scenario 
(the correlation length in Fig.\ \ref{fig:nu} further suggests the equal amplitudes $\xi_-=\xi_+$). 
Furthermore, these exponents are consistent, within numerical accuracy, with the constraints \eqref{eq:nu_beta_eta} expected for the Gaussian universality class. 
The TLL parameter $K_+$ in the symmetric channel discussed in Sec.\ \ref{sec:EFT} is equal to the exponent $\eta_{\text{N\'eel/SD}}$, whose value is approximately $0.6$ in Table \ref{tab:critical_exponents}. 
Thus, the higher-frequency cosine potential with the scaling dimension $8K_+$ discussed in Sec.\ \ref{sec:bos_phases} is expected to be irrelevant, lending further support to the scenario of the Gaussian transition. 
The critical exponents in Table \ref{tab:critical_exponents} do not detectably change along the phase boundary while they are allowed to do so for the Gaussian class.

In this work, we
have focused 
on the two-leg ladder system.
In order to reveal the relationship between 
our 1D $J$-$Q$-like 
model and the 2D $J$-$Q$ model, 
it would be interesting 
to investigate the dependence on the number of legs 
by studying, e.g., three- and four-leg ladder systems. 
It would also be interesting to study the N\'eel-SD transition in the 2D $J$-$Q$ model with a positive coefficient $-Q>0$ in the 
four-spin interaction, 
for which the QMC suffers from a sign problem but tensor network algorithms could be applicable.


\begin{acknowledgments}
The authors would like to thank 
A.~Furusaki, S.~Iino, R.~K.~Kaul, H.~Kohshiro, K.~Tamai, and L.~Vanderstraeten
for stimulating discussions.
This research was supported by JSPS KAKENHI Grant Number JP18K03446, JP19H01809 and JP20K03780,
and by MEXT as ``Priority Issue on Post-K computer'' 
(Creation of New Functional Devices and High-Performance Materials to Support Next-Generation Industries)
and ``Exploratory Challenge on Post-K computer'' (Challenge of Basic Science---Exploring Extremes through Multi-Physics and Multi-Scale Simulations).
The numerical computations were performed on computers at the Supercomputer Center, 
the Institute for Solid State Physics (ISSP), the University of Tokyo. 
\end{acknowledgments}

\appendix


\section{ 
Numerical analysis of the 
RS-N\'{e}el transition \label{subsec:RSNeel} }
In this appendix, 
we describe our numerical results on 
the RS-N\'{e}el transition. 
We fix $J = \Jperp = 1$ and study the transition at $\Delta = 1.1$ as a representative case. 
The procedure is similar to our analysis of the N\'eel-SD transition in Sec.\ \ref{sec:Results}. 

By analyzing the correlation length and the entanglement entropy, 
we obtain the critical point 
$J_{4,c} \simeq 0.666$
and the central charge $c \simeq 0.48$.
Around the transition point, the correlation length shows a consistent growth as a function of $\chi$, which indicates a continuous nature of the transition. 
The ciritical exponents are 
estimated to be 
$\beta \simeq 0.12$, $\nu_{-} \simeq 1.0$, $\nu_{+} \simeq 1.0$, and $\eta \simeq 0.26$.

The above numerical results are consistent with the (1+1)-dimensional Ising universality class 
with $c=1/2$, $\beta=1/8$, $\nu=1$, and $\eta=1/4$. 
We note that the occurrence of an Ising transition is also suggested by the field-theoretical analysis in Sec.\ \ref{sec:EFT}. 
We have obtained similar 
results on other points along the boundary between the RS and N\'eel phases.

\section{ Numerical analysis of the RS-SD transition in the isotropic case
\label{app:RSSD} }
In this appendix,
we describe our numerical results on the RS-SD transition 
in the isotropic case $\Delta=1$ with $J=\Jperp=1$
and $J_4\simeq1.19$. 
The previous study based on exact diagonalization \cite{PhysRevB.80.014426} has suggested that this transition is described by the $\text{SU(2)}_2$ WZW model with $c=3/2$.
Below we 
show that our VUMPS results are consistent with the previous 
study. 

\begin{figure}[t]
\includegraphics[width=90mm]{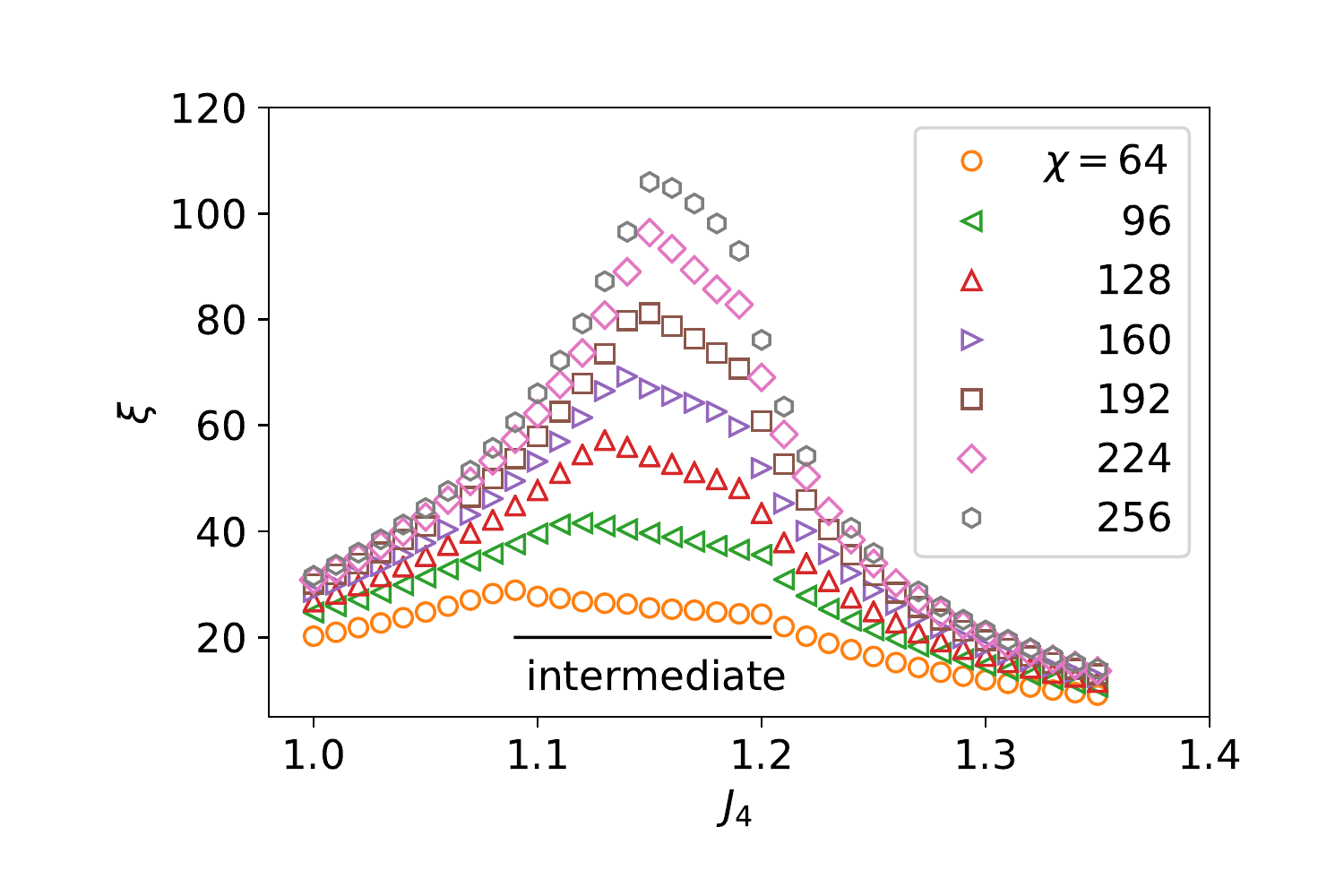}
\caption{
Correlation length $\xi$ as a function of $J_4$ around the 
expected 
RS-SD transition 
in the isotropic case $\Delta=1$. 
The correlation length grows consistently with an increase in $\chi$,
but the correlation length 
for fixed $\chi$ exhibits two kinks instead of a single sharp peak. 
The intermediate region between the two kinks tends to shrink with an increase in $\chi$, 
as we analyze more closely in Fig.\ \ref{fig:RSSDintermediate}.}

\label{fig:RSSDcorr}
\end{figure}

\begin{figure}[t]
\includegraphics[width=90mm]{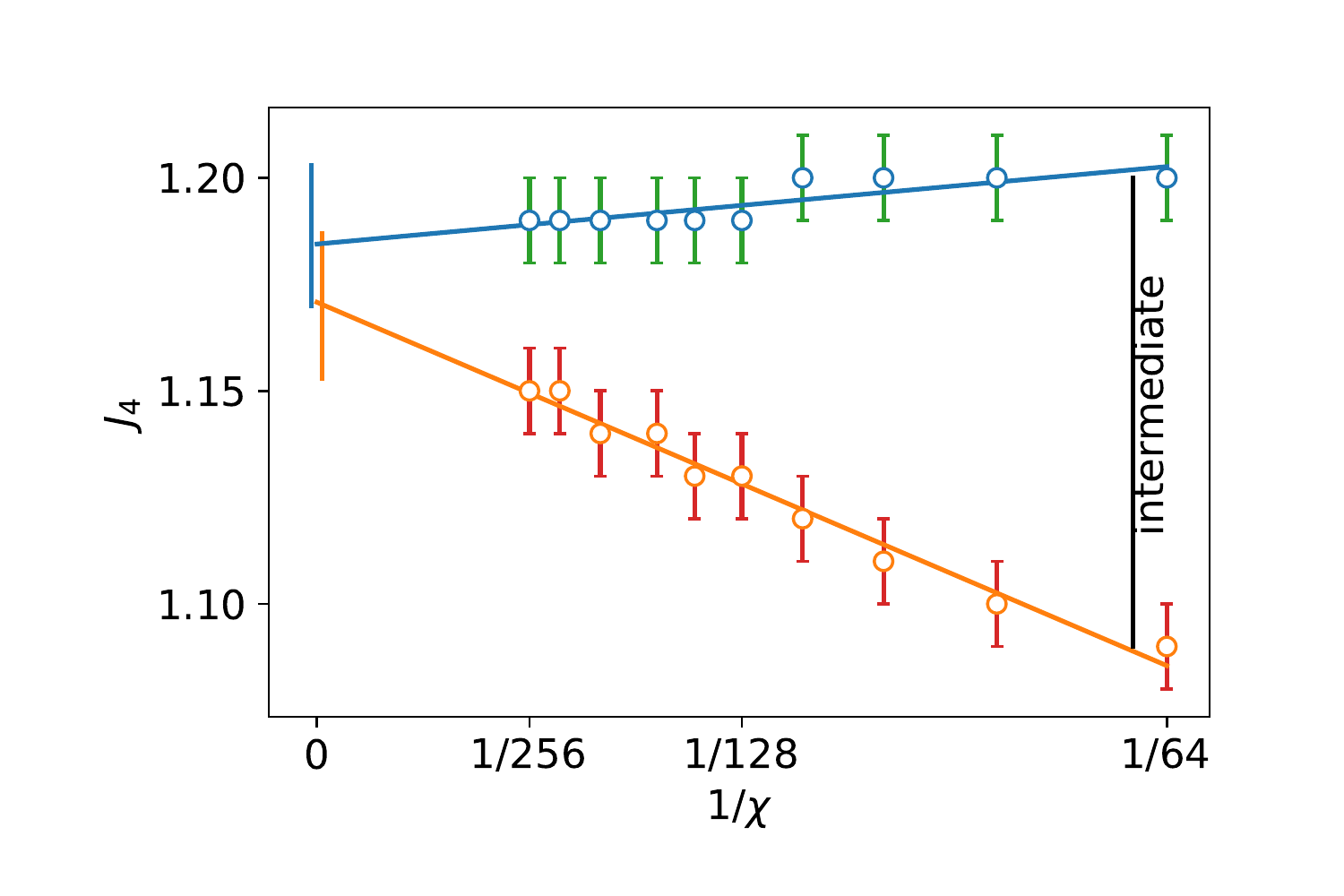}
\caption{
Dependence of the intermediate region found in Fig.\ \ref{fig:RSSDcorr} on the bond dimension $\chi$. The extrapolation to infinite $\chi$ is consistent with the vanishing of this region within numerical accuracy.
}
\label{fig:RSSDintermediate}
\end{figure}

\begin{figure}[t]
\includegraphics[width=90mm]{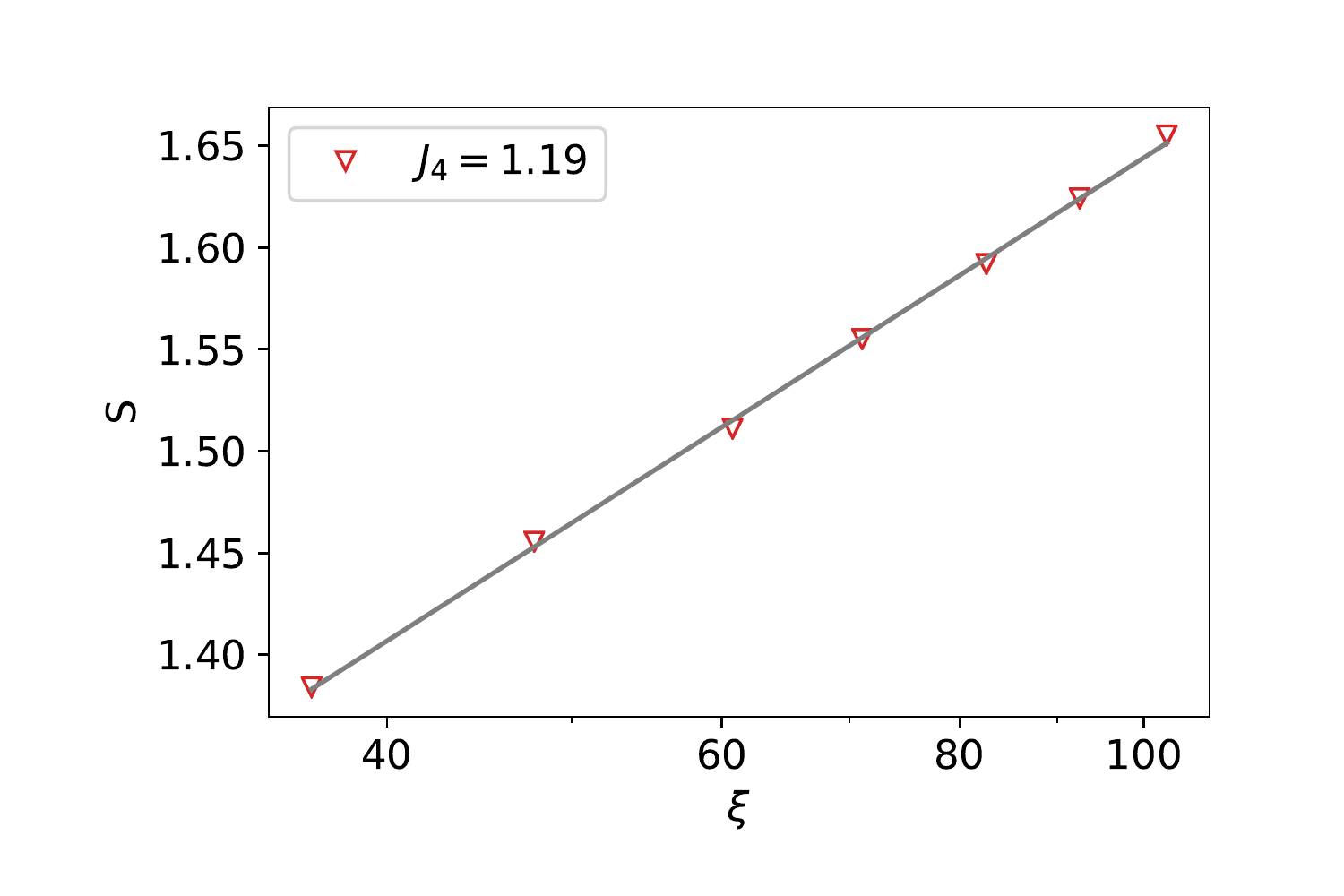}
\caption{
Entanglement entropy $S(\chi)$ versus the correlation
length $\xi(\chi)$ for the bond dimension $\chi = 96,128,160,192,224,256,280$.
These are calculated 
in the isotropic case $\Delta=1$ with $J=\Jperp=1$ and $J_4=1.19$, around which the transition between the RS and SD phases is expected to occur.
A logarithmic scale is used for the horizontal axis.
The gray straight line shows the CFT formula Eq.\ (\ref{eq:cardy}) with $c\simeq 1.55$ and $S_0 \simeq 0.45$,
which is the result of a fitting to 
the numerical data. 
}
\label{fig:RSSDCC}
\end{figure}

The correlation length 
$\xi$ calculated by VUMPS around the expected 
RS-SD transition exhibits two kinks instead of a single sharp peak
as shown in Fig. \ref{fig:RSSDcorr}. 
In the intermediate region between the two kinks, the N\'eel order parameter $\expval{\mathcal{O}_{\text{N\'eel}}}$ has been found to be nonzero (not shown). 
As the antiferromagnetic order that spontaneously breaks the SU$(2)$ symmetry is not likely to appear in the present ladder system (because of the theorem in Ref.\ \cite{Momoi1996}), 
the intermediate region with $\expval{\mathcal{O}_{\text{N\'eel}}}\ne 0$ can be considered as an artifact of the present numerical method whose accuracy is controlled by the bond dimension $\chi$. 
With an increase in $\chi$, the intermediate region indeed tends to shrink, and the correlation length $\xi$ around this region continues to increase as seen in Fig. \ref{fig:RSSDcorr}. 
The extrapolation of the intermediate region to infinite $\chi$ shown in Fig.\ \ref{fig:RSSDintermediate} is consistent with the vanishing of this region within numerical accuracy. 
These results are indicative of a direct continuous transition between the RS and SD phases in the isotropic case. 
Assuming this is true, we obtain the central charge $c\simeq1.55$ by fitting Eq.\ \eqref{eq:cardy} to the data of the entanglement entropy versus the correlation length, as shown in Fig.\ \ref{fig:RSSDCC}.


\section{Estimation of the correlation length exponents 
\label{app:corr}}

\begin{table}[b]
\caption{\label{tab:critical_exponents_nu}
Numerical estimates of the
critical exponents
$\nu_\pm$ for the N\'eel-SD transition with different values of the assumed transition point $J_{4,c}$. 
The case of $J_{4,c}=1.951$ corresponds to Fig.\ \ref{fig:nu}. 
}
\label{tab:expnu}
\begin{ruledtabular}
\begin{tabular}{ccc}
\multicolumn{1}{c}{assumed $J_{4,c}$} & 
\multicolumn{1}{c}{$\nu_{-}$} & 
\multicolumn{1}{c}{$\nu_{+}$} \\
\hline
$1.950$ & $1.10(8)$ & $1.27(14) $\\
$1.951$ & $1.12(9)$ & $1.25(14)$\\
$1.952$ & $1.15(9)$ & $1.24(14)$ 
\end{tabular}
\end{ruledtabular}
\end{table}

In this appendix, we describe how we estimate 
the correlation length critical 
exponents $\nu_\pm$ and their error ranges. 
First, for 
given $\Delta$, we extrapolate the correlation length to infinite $\chi$ at each $J_4$.
Second, we determine the region that is 
used for the fitting.
Finally, we fit 
the scaling form Eq.\ \eqref{eq:nu} to the selected data points. 
As an example, we explain the procedure by taking the N\'{e}el-SD transition 
for $\Delta=1.2$ as shown in Fig.\ \ref{fig:nu}.

As discussed in Sec.\ 
\ref{subsec:NSD},
we 
tend to 
underestimate the correlation length 
when we use the transfer matrix with the finite bond dimension $\chi$.
Since this underestimation is not limited to 
the critical point,
we need 
to perform an appropriate extrapolation to infinite $\chi$ 
at each point.
Several extrapolation 
methods for 1D systems are known 
\cite{PhysRevX.8.041033}. 
In these methods, 
$1/\xi$ is regarded as a linear function of $1/\chi$ or $|\lambda_2 - \lambda_3|$,
where $\lambda_2$ and $\lambda_3$ are the 
second and third largest absolute eigenvalues 
of the transfer matrix.
However, we cannot use these methods directly 
because our model is a ladder system 
and its transfer matrix has 
a 
different eigenvalue distribution from 
those of 1D 
chains.
We 
instead 
use the linear scaling of the correlation length $\xi(\chi)$ 
with 
$1/\sqrt{\chi}$.
The reason why we use $\sqrt{\chi}$ instead of $\chi$ is as follows.
When we have two independent chains each of which is described by 
a uniform MPS 
with 
the bond dimension 
$\sqrt{\chi}$,
where we assume that $\chi$ is a square number,
the whole system is a two-leg ladder system without 
inter-chain couplings 
and is described by
a uniform MPS
with $\chi$.
In this case, the relationship between $\xi$ and $\chi$ in 
the two-leg ladder 
is the same as the relationship between $\xi$ and $\sqrt{\chi}$ in 
1D 
chains.
We assume that the above discussion is 
applicable 
to our system
having inter-chain couplings. 
This method is not suitable near the critical point 
because the correlation length at the critical point 
has been predicted to follow 
$\xi \propto \chi^\kappa$, 
where $\kappa =6/\left[ c\left( \sqrt{12/c}+1 \right) \right]$ with $c$ being the central charge  \cite{PhysRevLett.102.255701}
\footnote{We indeed found that our data at the N\'eel-SD transition point followed this power-law scaling but with $\kappa\simeq 1.2$ instead of $\kappa =6/(\sqrt{12}+1)\simeq1.344$ for $c=1$. 
The disagreement in $\kappa$ is possibly due to the interplay between the symmetric and antisymmetric channels for a finite bond dimension $\chi$. }. 
%
%
To estimate 
a possible error in $\nu_\pm$, 
we assume that the extrapolated correlation length $\xi(\chi\rightarrow \infty)$ has 
its error at worst $5\%$ in our two-leg ladder model 
\footnote{This error estimation is expected to be sufficient for the following reason. 
We performed extrapolation assuming the linear scaling of $\xi(\chi)$ with $1/\chi$ for the transverse-field Ising chain, 
and the extrapolated correlation length $\xi(\chi\to\infty)$ was off the exact solution at worst 5 \%. 
This indicates that if we had two decoupled copies of the transverse-field Ising chains and described them using a single uniform MPS, 
extrapolation based on the linear scaling of $\xi(\chi)$ with $1/\sqrt{\chi}$ would result in the same error.}.
Note that the error is not a statistical one but a systematic one.


Next, we 
determine the fitting region.
At $\Delta = 1.2$, we 
first 
fix the critical point at $J_{4,c} = 1.951$,
which is determined in Sec. 
\ref{subsec:NSD}. 
As we discussed above, we cannot use the data points near the critical point
because the estimated correlation length has larger error 
there. 
Additionally, we cannot use the data points that are far from the critical point.
This is because they do not follow the power-law scaling as shown in Fig.\ \ref{fig:nu}.  
By fitting 
the scaling form Eq.\ \eqref{eq:nu} to the selected data points, 
we obtain
$\nu_{-} = 1.12\pm0.09$ and $\nu_{+} = 1.25\pm0.14$.
However, 
the errors in these results 
do not include 
the effect of a possible error 
in the critical point
and would be underestimated.
To estimate the error more reliably, we 
have to 
take into account the error 
in 
the critical point $J_{4,c}$.
Thus, we repeat the same procedure assuming 
that 
the critical point is $J_{4,c} = 1.950$ and $J_{4,c} = 1.952$.
The results are shown in Table \ref{tab:critical_exponents_nu}.
Thus, we conclude that 
$\nu_{-} =  1.13(11)$ and $\nu_{+} =  1.26(16)$.

\section{Estimation of the order parameter exponents 
\label{app:beta}}

\begin{table}[b]
\caption{\label{tab:critical_exponents_beta}
Numerical estimates of the
critical exponents 
$\beta_{\text{N\'eel/SD}}$ for the N\'eel-SD transition with different values of the assumed transition point $J_{4,c}$. 
The case of $J_{4,c}=1.951$ corresponds to Fig.\ \ref{fig:beta}.
}
\begin{ruledtabular}
\begin{tabular}{ccc}
\multicolumn{1}{c}{assumed $J_{4,c}$} & 
\multicolumn{1}{c}{$\beta_{\text{N\'{e}el}}$} & 
\multicolumn{1}{c}{$\beta_{\text{SD}}$} \\
\hline
$1.950$ & $0.329(18)$ &  $0.352(18)$\\
$1.951$ & $0.338(18)$ &  $0.342(17)$\\
$1.952$ & $0.347(19)$ &  $0.331(17)$
\end{tabular}
\end{ruledtabular}
\end{table}

In this appendix, we describe how we 
estimate the order parameter critical
exponents $\beta_{\text{N\'eel/SD}}$ and their error ranges. 
The procedure is essentially the same as the estimation of the correlation length 
exponents $\nu_\pm$ 
in Appendix \ref{app:corr}.
Here, we fit 
the scaling form Eq.\ \eqref{eq:beta} to the selected data points.

As shown in Fig. \ref{fig:beta}, 
we observe a kink at 
$ \Delta J_4 \sim 0.008$
for both order parameters.
We 
tend to 
overestimate the order parameters near the critical point
 ($\Delta J_4 < 0.008$)
because $\chi$ is not large enough.
To estimate the gap between the order
parameter 
$\expval{\mathcal{O}(\chi=256)}$ and the exact one,
we extrapolate the order parameter by fitting the data points 
with 
a function
\begin{eqnarray}
f(\chi) = a + b \, \qty(\frac{1}{\chi})^c,
\end{eqnarray}
where $a$, $b$, and $c>1$ are fitting parameters.
We confirm that $f(\chi\rightarrow\infty)=a$ almost always underestimates the order parameter
by comparing the extrapolated 
order parameter 
with the exact one in the
transverse-field Ising chain.
Thus, we expect that the exact order parameter is
between $\expval{\mathcal{O}(\chi=256)}$ and $f(\chi\rightarrow\infty)$.
The error of the order parameter is estimated to be smaller than 
\begin{eqnarray}
\Delta\expval{\mathcal{O}(\chi=256)}
:= 
\frac{|\expval{\mathcal{O}(\chi=256)}-f(\chi\rightarrow\infty)|}{|\expval{\mathcal{O}(\chi=256)}|}. 
\nonumber \\
\end{eqnarray}

Next, we need to determine the fitting region. 
We fix the critical point at $J_{4,c} = 1.951$,
which is determined in Sec.\ 
\ref{subsec:NSD},
We choose the data points with $\Delta\expval{\mathcal{O}(\chi=256)} < 10^{-2}$
and discard the data points that are far from the critical point
because they do not follow the power-law scaling Eq.\ \eqref{eq:beta} as shown in Fig.\ \ref{fig:beta}.
By fitting 
the scaling form Eq.\ \eqref{eq:beta} to the selected data points, 
we obtain $\beta_{\text{N\'{e}el}}= 0.338(18)$ and $\beta_{\text{SD}} = 0.342(17)$
with the estimated critical point $J_{4,c}=1.951$,
assuming that all data points have $1\%$ systematic error. 
In the same way,
we obtain the critical exponents in Table. \ref{tab:critical_exponents_beta}.
As a result, we obtain $\beta_{\text{N\'{e}el}} = 0.34(3)$ and $\beta_{\text{SD}} = 0.34(3)$.

\bibliography{apssamp}

\end{document}